\documentclass[journal=jctce,manuscript=article,layout=traditional]{achemso}

\usepackage{graphicx}
\usepackage{epsfig} 
\usepackage{amssymb}
\usepackage{amsmath}
\usepackage{color}
\usepackage{tabularx}
\usepackage{url}
\usepackage{hyperref}
\usepackage{capt-of}
\usepackage{siunitx}
\SectionNumbersOn

\newcolumntype{C}[1]{>{\centering\let\newline\\\arraybackslash\hspace{0pt}}m{#1}}
\newcommand\T{\rule{0pt}{3ex}}       
\newcommand\B{\rule[-1.2ex]{0pt}{0pt}} 

\allowdisplaybreaks

\author{Nicola Colonna}
\affiliation[epfl]{Theory and Simulation of Materials (THEOS) \\ and National Centre for Computational Design and Discovery of Novel Materials (MARVEL),
\'Ecole Polytechnique F\' ed\'erale de Lausanne, 1015 Lausanne, Switzerland} 
\email{nicola.colonna@epfl.ch}
\author{Ngoc Linh Nguyen}
\affiliation[epfl]{Theory and Simulation of Materials (THEOS) \\ and National Centre for Computational Design and Discovery of Novel Materials (MARVEL), \'Ecole Polytechnique F\' ed\'erale de Lausanne, 1015 Lausanne, Switzerland}
\alsoaffiliation[uchicago]{Institute for Molecular Engineering, University of Chicago, 5640 South Ellis Avenue, Chicago, Illinois 60637, USA}
\author{Andrea Ferretti}
\affiliation[cnr]{Centro S3, CNR-Istituto Nanoscienze, 41125 Modena, Italy} 
\author{Nicola Marzari}
\affiliation[epfl]{Theory and Simulation of Materials (THEOS) \\ and National Centre for Computational Design and Discovery of Novel Materials (MARVEL),
\'Ecole Polytechnique F\' ed\'erale de Lausanne, 1015 Lausanne, Switzerland} 

\title{Koopmans-compliant functionals and potentials and their application to the GW100 test-set}

\date{\today} 

\begin{document}

\begin{abstract}
Koopmans-compliant (KC) functionals have been shown to provide accurate spectral properties through 
a generalized condition of piece-wise linearity of the total energy as a function of the fractional 
addition/removal of an electron to/from any orbital. We analyze the performance of different KC functionals
on the GW100 test-set, comparing the ionization potentials (as opposite of the energy of the highest occupied orbital)
of these 100 molecules to those obtained from CCSD(T) total energy differences, and experimental results, finding 
excellent agreement with a mean absolute error of 0.20 eV for the KIPZ functional, that is state-of-the-art for both DFT-based
calculations and many-body perturbation theory.
We highlight similarities and differences between KC functionals and  other electronic-structure approaches,
such as dielectric-dependent hybrid functionals and G$_0$W$_0$, both from a theoretical and from a practical point of view,
arguing that Koopmans-compliant potentials can be considered as a local and orbital-dependent counterpart to the electronic
GW self-energy, albeit already including approximate vertex corrections.
\end{abstract}

\section{Introduction}
Computational spectroscopy is a valuable tool to support and complement experiments and  
drive, nowadays, the discovery of novel materials for diverse applications~\cite{yu_identification_2012, 
castelli_computational_2012, jain_computational_2016, pham_electronic_2017}.
Although reasonably accurate spectral properties can be obtained using quantum chemistry wave-function methods
or many-body perturbation theory ~\cite{onida_electronic_2002} (MBPT), these approaches 
have an unfavorable scaling with the size of the system often constraining one to
study only relatively small or simple systems. For this reason faster approaches based on 
density-functional theory (DFT), Hartree-Fock, or a mixture of the two,
as in hybrid functionals, are often used. $\Delta$ 
self-consistent ``$\Delta$SCF'' calculations based on standard density-functional
approximations (DFAs) are also quite accurate for small-size systems, but fails in the 
solid-state limit; single-particle energies from Hartree-Fock are protected by  
Koopmans' theorem and can be interpreted as energy removals or additions, but lack 
relaxation effects, usually leading to an overestimation of the first ionization potential (IP), an 
underestimation of the electron affinity (EA) and consequently to  overestimation of the fundamental gap. 
The exact exchange-correlation (xc) energy functional of Kohn-Sham (KS) density functional theory
would provide exactly the first ionization potential of a system of interacting electrons, as
the opposite of the energy of the highest occupied molecular orbital (HOMO) of the auxiliary KS system~\cite{perdew_density-functional_1982,almbladh_exact_1985} (in addition, of course,
of providing it as a total energy difference between the neutral system with $N$ electrons and
the ionized one with $N-1$ electrons). 
However ($i$) KS energies other than the HOMO have no obvious connection with charged excitations 
of the real system (see e.g. Ref.~\citenum{chong_interpretation_2002} for an in depth discussion about the
connection between KS eigenvalues and vertical ionization potentials), and ($ii$) not even the 
first ionization potential is correctly reproduced when standard local or semi-local DFAs are used.
This latter failure can be understood in terms of the deviation from the expected piece-wise linearity
(PWL) of the exact energy functional as a function of the number of particles, first discussed 
by Perdew-Parr-Levy-Balduz~\cite{perdew_density-functional_1982} (PPLB).
Diverse xc functional developments~\cite{cococcioni_linear_2005, kulik_density_2006, dabo_non-koopmans_2009, 
dabo_koopmans_2010, stein_fundamental_2010, zheng_improving_2011, kraisler_piecewise_2013, gorling_exchange-correlation_2015,
li_local_2015, kronik_excitation_2012, refaely-abramson_quasiparticle_2012,li_localized_2018} are indeed guided by such a fundamental condition,
with the aim of correcting the spurious self-interaction error present in almost all
DFAs and get accurate prediction of frontier orbital energies from single-particle energies. 

In this respect, Koopmans' compliant (KC) functionals have been introduced~\cite{dabo_non-koopmans_2009,dabo_koopmans_2010,borghi_koopmans-compliant_2014} to purify standard
density-functional approximations from self-interaction errors deriving from the lack of PWL. In this formulation, a generalized PWL
condition is imposed to the total energy as a function of the fractional occupation of
{\it any } orbital in the system, thus extending the standard PPLB linearity condition (valid only
for the highest occupied state) to the entire electronic manifold. This ansatz results in a beyond-DFT
orbital-density dependent functionals with enough flexibility to correctly reproduce 
both ground state and spectral properties. 
Infact, while the ground state energies are very close or identical to  those of the starting functional~\cite{borghi_koopmans-compliant_2014},
we argued~\cite{ferretti_bridging_2014} that for the spectral properties the orbital-dependent KC potentials
act as a quasiparticle approximation to the spectral potential~\cite{gatti_transforming_2007,vanzini_dynamical_2017,vanzini_spectroscopy_2018}, 
i.e. the local and frequency-dependent potential sufficient to correctly describe the local spectral density $\rho(\mathbf{r},\omega)$. 
To further support this picture we present in this work a detailed analysis of the KC orbital-dependent 
potentials and establish a connection with more complex electronic-structure methods, and in
particular with the well-known GW approximation~\cite{hedin_new_1965} in the framework of many-body perturbation theory. 
We provide additional evidence that, in a localized representation of the electronic manifold, 
the orbital-dependent KC potentials provide simplified but yet accurate approximations to the 
non-local and frequency-dependent self-energy.
We finally analyze the performance of the KC functionals on the GW100 test set~\cite{van_setten_gw100_2015};
comparing our results against state-of-the-art GW calculations, accurate quantum chemistry methods, and experiments
finding overall an excellent agreement, even more remarkable as obtained with a functional theory of the orbital densities.

\section{Koopmans-compliant functionals}
\label{sec:KC_func}
In this Section we review the basic features of KC functionals and we refer to Refs.~\citenum{borghi_koopmans-compliant_2014,
nguyen_koopmans-compliant_2017, colonna_screening_2018} for an extensive discussion of the approach,
that is based on the three concepts of linearization, screening, and localization.

{\it Linearization.} The basic ansatz behind the
 KC functionals is to enforce a generalized criterion of piece-wise linearity to the total energy as a function of 
the fractional occupation of any orbital in the system. This is a generalization of the well-know PWL of the 
total energy as a function of the number of particles~\cite{perdew_density-functional_1982} (or, equivalently, as 
a function of the fractional occupation of the highest occupied state) to the entire electronic manifold. 
This is achieved in two steps: starting from any non-linear density functional $E^{\rm DFT}$ 
and for each orbital $\phi_{i\sigma}$
first (i) the ``bare'' or ``unscreened'' Koopmans corrections $\{\Pi^u_{i\sigma}\}$ are applied to enforce linearity as a 
function of each spin-orbital occupation $f_{i\sigma}$ in a {\it frozen orbital picture}, i.e. assuming that all other electrons do not 
respond when a particle is added/removed from the system. Then (ii) relaxation effects are 
captured by orbital-dependent screening parameters $\{\alpha_{i\sigma}\}$ for the bare corrections,
leading to the general form of the KC functionals as:
\begin{equation}
 E^{\rm KC} = E^{\rm DFT} + \sum_{i\sigma} \alpha_{i\sigma} \Pi^u_{i\sigma}.
 \label{eq:KC}
\end{equation}
The corrections $\{\Pi_{i\sigma}^u\}$ remove, in a frozen-orbital picture and for each orbital $\phi_{i\sigma}$, the ``Slater'' non-linear 
behavior of the underlying functional as a function of the occupation $f_{i\sigma}$ of the orbital at hand, and 
replace it with a linear Koopmans term; that is
\begin{align}
 \Pi_{i\sigma}^u(f_{i\sigma}) = & -\int_0^{f_{i\sigma}} \langle \phi_{i\sigma} | H^{\rm DFT}(s) | \phi_{i\sigma} \rangle \, ds +  
                f_{i\sigma} \eta_{i\sigma} \\
              = &-\text{Slater}+\text{Koopmans}
 \label{eq:uPi_KC}
\end{align}
where $H^{\rm DFT}(s)$ is the KS-DFT Hamiltonian calculated at a 
density where the orbital $\phi_{i\sigma}$ has occupation $s$, and $\eta_{i\sigma}$ is the slope of the linear Koopmans
term. The slope $\eta_{i\sigma}$ can be chosen either following Slater's intuition (i.e. choosing $\eta_{i\sigma}$ as the orbital 
energy at occupation $1/2$), or enforcing in a functional form the $\Delta$SCF concept, i.e. choosing 
the slope as the total energy difference between two adjacent points at integer occupations ($f_{i\sigma}=0$ and $f_{i\sigma}=1$).
In the latter case one obtains the KI (``I'' stands for integral) correction, that reads
\begin{align}
 \Pi^{\rm uKI}_{i\sigma} = & - \int_0^{f_{i\sigma}} \langle \phi_{i\sigma} | H^{\rm DFT}(s) | \phi_{i\sigma} \rangle \, ds +
                       f_{i\sigma}\int_0^{1} \langle \phi_{i\sigma} | H^{\rm DFT}(s) | \phi_{i\sigma} \rangle \, ds = \nonumber \\
                 = & - \Big( E^{\rm DFT}_{\rm Hxc}[\rho] - E^{\rm DFT}_{\rm Hxc}[\rho-\rho_{i\sigma}]\Big) 
                     + f_{i\sigma}\Big( E^{\rm DFT}_{\rm Hxc}[\rho-\rho_{i\sigma}+n_{i\sigma}] - E^{\rm DFT}_{\rm Hxc}[\rho-\rho_{i\sigma}]\Big)
  \label{eq:uPi_KI}
\end{align}
where $E^{\rm DFT}_{\rm Hxc}$ is the Hartree and exchange-and-correlation energy associated to the underlying 
functional, $\rho_{i\sigma}(\mathbf{r}) = f_{i\sigma}|\phi_{i\sigma}(\mathbf{r})|^2 = f_{i\sigma} n_{i\sigma}(\mathbf{r})$
is the orbital density at occupation $f_{i\sigma}$, $n_{i\sigma}(\mathbf{r})$ is the orbital density at occupation 1,
and $\rho(\mathbf{r}) = \sum_{i\sigma} \rho_{i\sigma}(\mathbf{r})$ is the total charge density. 
From the definition above it can be seen that at integer occupations the $\Pi_{i\sigma}^{\rm uKI}$ correction 
vanishes and the KI functional becomes identical to its base functional, independently from the screening
coefficients. The KI functional thus preserves exactly the potential energy surface of the base functional
it started from; its value at fractional occupations is instead different and so are the derivatives everywhere, including
at integer occupations, and therefore the spectral properties.
If instead of the DFT energy difference we choose the Perdew-Zunger (PZ) self-interaction corrected~\cite{perdew_self-interaction_1981} (SIC) 
one, i.e. $\eta_{i\sigma}=\int_0^{1} \langle \phi_{i\sigma} | H_{i\sigma}^{\rm PZ}(s) | \phi_{i\sigma} \rangle \, ds$, we obtain the KIPZ
correction:
\begin{align}
 \Pi^{\rm uKIPZ}_{i\sigma} = & - \int_0^{f_{i\sigma}} \langle \phi_{i\sigma} | H^{\rm DFT}(s) | \phi_{i\sigma} \rangle \, ds +
                       f_{i\sigma} \int_0^{1} \langle \phi_{i\sigma} | H^{\rm PZ}_{i\sigma}(s) | \phi_{i\sigma} \rangle \, ds \nonumber \\
                 = & \phantom{-} \Pi_{i\sigma}^{\rm KI} - f_{i\sigma} E_{\rm Hxc}[n_{i\sigma}]
  \label{eq:uPi_KIPZ}
\end{align}
where $H^{\rm PZ}_{i\sigma}(s)=H^{\rm DFT}(s) - v^{\rm DFT}_{\sigma,\rm Hxc}[s|\phi_{i\sigma}(\mathbf{r})|^2]$ is the PZ-SIC Hamiltonian,
with $- v^{\rm DFT}_{\sigma,\rm Hxc}$ the PZ-SIC correction for the orbital $\phi_{i\sigma}$ occupied by $s$ electrons.
The KIPZ functional inherits from the PZ-SIC functional the important property of being one-electron self-interaction free (at least in its unscreened form, that is the exact one for 
one-electron systems),
while at the same time it preserves the linear behavior of the energy also in many-particle systems and thus is
also (approximately) free from the many-body self-interaction error.\cite{cohen_insights_2008}

{\it Screening.} If all $\alpha_{i\sigma}=1$ the energy functional in Eq.~(\ref{eq:KC}) is by construction linear in each orbital
occupation $f_{i\sigma}$ if one neglects orbitals relaxation. This is analogous to realizing a DFT equivalent of the ``restricted'' 
Koopmans theorem in Hartree-Fock theory, and it is not enough to guarantee the linearity of 
the functional in the general case. To include the response of the system to the ionization process
(change in the occupation) embodied in the $\Pi^{\rm KC}_{i\sigma}$ term, {\it screening parameters} $\{\alpha_{i\sigma}\}$ are introduced 
for each orbital. 
By definition the screening parameters transform the unrelaxed Koopmans' correction $\Pi^{\rm u}_{i\sigma}$ into the fully relaxed 
one $\Pi_{i\sigma}^{\rm r}$. The latter can be calculated by self-consistent finite differences~\cite{nguyen_koopmans-compliant_2017}, or 
by linear response~\cite{colonna_screening_2018} as if a tiny fraction of an electron were removed/added from/to a given orbital.
In this case a second-order Taylor expansion in the orbital occupations $f_{i\sigma}$ can be used to approximate 
the relaxed and unrelaxed KC corrections, and the screening coefficient $\alpha_{i\sigma}$ associated to each 
orbital $\phi_{i\sigma}$ can be calculated as~\cite{colonna_screening_2018}
\begin{align}
&\alpha_{i\sigma} = \frac{\Pi^{\rm r}_{i\sigma}}{\Pi_{i\sigma}^{\rm u}} \xrightarrow{\text{2$^{\circ}$ order}} 
 \frac{\left[d^2 E^{\rm app}/df_{i\sigma}^2\right]_{\rm relax}}{\left[d^2 E^{\rm app}/df_{i\sigma}^2\right]_{\rm unrelax}}
 =\frac{\langle n_{i\sigma} | \left[\tilde{\varepsilon}^{-1} f_{\rm Hxc}\right]^{\sigma\sigma}| n_{i\sigma} \rangle }{\langle n_{i\sigma} | f^{\sigma\sigma}_{\rm Hxc}| n_{i\sigma} \rangle}
 \label{eq:alpha_1}
\end{align}
where
\begin{align}
 & \Pi_{i\sigma}^{\rm u} = \frac{1}{2}f_{i\sigma}(1-f_{i\sigma}) \left.\frac{d^2E^{\rm app}}{df_{i\sigma}^2}\right|_{\rm unrelax} + O(f_{i\sigma}^3), \label{eq:alpha_2}
 \\
 & \Pi_{i\sigma}^{\rm r} = \frac{1}{2}f_{i\sigma}(1-f_{i\sigma}) \left.\frac{d^2E^{\rm app}}{df_{i\sigma}^2}\right|_{\rm relax} + O(f_{i\sigma}^3) \label{eq:alpha_3}
 \\
 & \tilde{\varepsilon}^{-1} = 1 + f_{\rm Hxc}\chi. \label{eq:alpha_4}
\end{align}
Here $\chi(\mathbf{r},\mathbf{r}')$,  $\tilde{\varepsilon}(\mathbf{r},\mathbf{r}')$ and 
$f^{\sigma\sigma'}_{\rm Hxc}(\mathbf{r,\mathbf{r}'})=\frac{\delta E^{\rm app}_{\rm Hxc}}{\delta \rho_{\sigma}(\mathbf{r})\delta \rho_{\sigma'}(\mathbf{r}')}$ 
are the density-density response function, the microscopic dielectric matrix 
and the Hxc kernel, respectively, evaluated at the underlying approximate (app) functional~\cite{colonna_screening_2018}.
The latter would be a standard density functional 
for KI and the PZ-SIC functional for KIPZ~\cite{fhxc_pzsic}. While in the former case an efficient
implementation based on the linear-response technique of density-functional perturbation theory 
can be used~\cite{colonna_screening_2018}, for the KIPZ functional a less elegant although 
straightforward finite difference approach is needed~\cite{nguyen_koopmans-compliant_2017}
since analytical derivatives and linear-response techniques for the PZ functional are not
available in standard DFT codes~\cite{giannozzi_advanced_2017,gonze_recent_2016}. 

{\it Localization.} The KC corrections in Eqs.~(\ref{eq:uPi_KI}) and (\ref{eq:uPi_KIPZ}) depend on each orbital density and the 
KC functional is therefore orbital-density dependent. At variance with density dependent functionals, orbital-density
dependent functionals can break the invariance of the total energy against a unitary rotation of the occupied manifold.~\footnote{ 
This is the case for the PZ-SIC and KIPZ functional, while for the KI functional at integer occupations the energy
corrections vanish and the KI energy remains identical to the one of the underlying density functional~\cite{borghi_koopmans-compliant_2014}.} 
For such functionals the {\it variational orbitals} that minimize the functional are different from the 
eigenstates or {\it canonical orbitals} that diagonalize the orbital-density dependent Hamiltonian (the Lagrange multiplier matrix), as
discussed e.g in Refs.~\citenum{borghi_koopmans-compliant_2014},~\citenum{ferretti_bridging_2014} and Refs.~\citenum{nguyen_first-principles_2015,
lehtola_variational_2014,vydrov_tests_2007,pederson_localdensity_1984,stengel_self-interaction_2008,hofmann_using_2012} .
Such duality is an important feature of any orbital-density dependent scheme. In general the variational orbitals 
exploit the additional freedom of unitary mixing to become localized~\cite{borghi_koopmans-compliant_2014,pederson_localized_1988} (and 
similar to Wannier functions~\cite{edmiston_localized_1963, boys_construction_1960, foster_canonical_1960,marzari_maximally_2012})
in order to further lower the total energy~\cite{ki_invariance}. On the other side the canonical orbitals and energies are 
the analogous of the single-particle KS-DFT or Hartree-Fock states and their energies are directly connected to those of 
Dyson orbitals and quasiparticle energies accessible, e.g., from photoemission experiments.~\cite{ferretti_bridging_2014, nguyen_first-principles_2015}

\subsection{KC as a simplified electronic self-energy}\label{sec:cohsex}

To establish a connection with common electronic-structure approaches, we 
discuss here in some detail the actual form of the KC energy and potential corrections.
To simplify the discussion we resort again to a Taylor expansion of the functionals in the 
orbital occupations. Following Ref.~\citenum{colonna_screening_2018} we can write 
the fully relaxed KI and KIPZ corrections up to second order in the occupations as
\begin{eqnarray}
 \Pi_{i\sigma}^{\rm rKI(2)}   &=& \frac{1}{2}f_{i\sigma}(1-f_{i\sigma})\frac{d^2E^{\rm DFT}}{df_{i\sigma}^2} = \frac{1}{2}f_{i\sigma}(1-f_{i\sigma}) \langle n_{i\sigma} | \mathcal{F}^{\sigma\sigma}_{\rm Hxc} | n_{i\sigma} \rangle,  \\
 \Pi_{i\sigma}^{\rm rKIPZ(2)} &=& \Pi_{i\sigma}^{\rm rKI (2)} -f_{i\sigma} E_{\rm Hxc}[n_{i\sigma}] = \frac{1}{2}f_{i\sigma}(1-f_{i\sigma}) \langle n_{i\sigma} | \mathcal{F}^{\sigma\sigma}_{\rm Hxc} | n_{i\sigma} \rangle - f_{i\sigma} E_{\rm Hxc}[n_{i\sigma}],
\end{eqnarray}
with $\mathcal{F}_{\rm Hxc}= \tilde{\varepsilon}^{-1} f_{\rm Hxc}$. 
The screened, additional, orbital dependent KC potential acting on the $i$-th electron is defined as the derivative of the relaxed 
KC energy $\Pi^{\rm KC} = \sum_k \Pi_k^{\rm rKC(2)} + O(f_k^3)$ with respect to the orbital density $\rho_{i\sigma}$.  
Adding also the xc contribution from the underlying DFT functional, the total orbital-density dependent xc  potentials read, up to the second order: 
\begin{eqnarray}
v_{i\sigma, \rm xc}^{\rm rKI(2)}(\mathbf{r})   &=& v^{\rm DFT}_{\sigma,\rm xc}(\mathbf{r}) 
                                             -\frac{1}{2} \langle n_{i\sigma} | \mathcal{F}^{\sigma\sigma}_{\rm Hxc}| n_{i\sigma} \rangle
                                              +(1-f_{i\sigma}) \int d\mathbf{r}'\,\mathcal{F}^{\sigma\sigma}_{\rm Hxc}(\mathbf{r},\mathbf{r}')n_{i\sigma}(\mathbf{r}'), \label{eq:ki_pot_2} \\
v_{i\sigma, \rm xc}^{\rm rKIPZ(2)}(\mathbf{r}) &=& v_{i\sigma, \rm xc}^{\rm rKI (2)}(\mathbf{r}) 
                                           -\left\{ E_{\rm Hxc}[n_{i\sigma}] - \int d\mathbf{r}'\, v_{\sigma,\rm Hxc}[n_{i\sigma}](\mathbf{r}')n_{i\sigma}(\mathbf{r}') +  v_{\sigma,\rm Hxc}[n_{i\sigma}](\mathbf{r}) \right\},
\label{eq:kipz_pot_2}                      %
\end{eqnarray}
where the orbital dependent Hamiltonian is defined as  $\hat{h}_{i\sigma}|\phi_{i\sigma}\rangle = \hat{h}_0|\phi_{i\sigma}\rangle + 
\hat{v}_{i\sigma, \rm xc}^{\rm rKC(2)}|\phi_{i\sigma}\rangle$, with $\hat{h}_0 = -\frac{1}{2}\nabla^2 + \hat{v}_{\rm ext}+\hat{v}_{\rm H}[\rho]$ 
the Hartree Hamiltonian. Note that in Eq.~(\ref{eq:ki_pot_2}) we neglected the variation of $\mathcal{F}^{\sigma\sigma}_{\rm Hxc}$ 
with respect to $\rho_{i\sigma}$ to stay within a second-order approximation. 

In the absence of relaxation effects, i.e. assuming $\tilde{\varepsilon}^{-1} = 1$, and 
neglecting the xc contribution in the underlying DFT, i.e. using a Hartree-only functional, 
the KIPZ correction provides a good approximation of the non-local Hartree-Fock (HF) exchange
operator when evaluated on a localized representation.~\cite{ferretti_bridging_2014} 
In fact, the HF self energy $\Sigma_{\rm x}$ is given, in terms of the occupied single particle spin-orbitals, as
\begin{eqnarray}
\Sigma_{\rm x}(\mathbf{x},\mathbf{x'})= -\sum^{\rm occ}_{i\sigma}\psi_{i\sigma}(\mathbf{x})\psi_{i\sigma}^*(\mathbf{x}')f_{\rm H}(\mathbf{r},\mathbf{r}')
\end{eqnarray}
where 
$\psi_{i\sigma}(\mathbf{x})$ are  spin-orbital wavefunctions and $\mathbf{x}=\{\mathbf{r},\xi\}$
is a composite variable for the spatial coordinate $\mathbf{r}$ and the spin coordinate $\xi$. In the present work 
we neglect relativistic effects such that 
$\psi_{i\sigma}(\mathbf{x})=\phi_{i\sigma}(\mathbf{r})\pi_{\sigma}(\xi)$ factorizes 
in the product of a spatial function $\phi_{i\sigma}(\mathbf{r})$ and a spin 
function $\pi_{\sigma}(\xi)$ (see e.g. Ref.~\citenum{szabo_modern_1996}).
Since $\Sigma_{\rm x}$ depends on the density matrix, it is invariant under unitary transformations of the occupied manifold and 
can be expressed on any equivalent representation of the occupied manifold. 
In a representation where the orbitals $\{\phi_{i\sigma}\}$ are as localized (non-overlapping) as possible, 
the off-diagonal contributions of the exchange operator can be neglected, and its matrix
elements become:
\begin{align}
 \langle {i\sigma} | \Sigma_{\rm x} | {j\sigma'} \rangle & =  -\sum_{k\sigma''}^{\rm occ} \int d\mathbf{x} \, \psi^*_{i\sigma}(\mathbf{x}) \psi_{k\sigma''}(\mathbf{x}) \int d\mathbf{x}' \frac{\psi^*_{k\sigma''}(\mathbf{x}')\psi_{j\sigma'}(\mathbf{x}')}{|\mathbf{r}-\mathbf{r}'|} \\
 & = -\sum_{k\sigma''}^{\rm occ} \int d\mathbf{r} \, \phi^*_{i\sigma}(\mathbf{r}) \phi_{k\sigma''}(\mathbf{r}) \delta_{\sigma\sigma''}\int d\mathbf{r}' \frac{\phi^*_{k\sigma''}(\mathbf{r}')\phi_{j\sigma'}(\mathbf{r}')}{|\mathbf{r}-\mathbf{r}'|} \delta_{\sigma''\sigma'} 
 \nonumber \\
 & \simeq  - \langle \phi_{i\sigma} | v_{\rm H}[n_{i\sigma}] | \phi_{i\sigma} \rangle \delta_{ij}\delta_{\sigma\sigma'},
\end{align}
where the Kronecker $\delta$ over the spin indices comes from the orthonormality of the spin functions.
On the other hand, the matrix elements of the Hartree-only KIPZ potential without screening follow from Eq.~\eqref{eq:kipz_pot_2}, when
any xc contribution is set to zero and $\tilde{\varepsilon}^{-1}=1$, and read
\begin{align}
 \langle {i\sigma} | v_{j\sigma', \rm xc}^{\rm uKIPZ(2)} | {j\sigma'} \rangle & \simeq \delta_{ij}\delta_{\sigma\sigma'} \left\{ \left(\frac{1}{2}-f_{i\sigma}\right) \langle n_{i\sigma} | f_{\rm H} | n_{i\sigma} \rangle - E_{\rm H}[n_{i\sigma}] \right\}
\label{eq:hkipz_u}                                              
\end{align}
that for occupied orbitals reduces to 
$-2E_{\rm H}[n_{i\sigma}]\delta_{ij}\delta_{\sigma\sigma'}
=- \langle \phi_{i\sigma} | v_{\rm H}[n_{i\sigma}] | \phi_{i\sigma} \rangle \delta_{ij}\delta_{\sigma\sigma'} $.

If we now turn on screening effects at the Hartree level, the RPA (statically) screened interaction $W=\varepsilon^{-1}_{\rm RPA} f_H$
appears in Eq.~(\ref{eq:hkipz_u}), instead of the bare Coulomb kernel, and the KIPZ matrix elements become:
\begin{equation}
\langle {i\sigma} | v_{j\sigma', \rm xc}^{\rm rKIPZ(2)} | {j\sigma'} \rangle \simeq \delta_{ij} \delta_{\sigma\sigma'} \left\{\left(\frac{1}{2}-f_{i\sigma}\right) \langle n_{i\sigma} | W | n_{i\sigma} \rangle - E_{\rm H}[n_{i\sigma}]\right\} .
\label{eq:hkipz_r}
\end{equation}
We now show that these matrix elements are similar to the static GW self-energy, known as 
Coulomb-hole plus screened-exchange (COHSEX) self-energy~\cite{hedin_new_1965,gygi_quasiparticle_1989,kang_enhanced_2010}:
\begin{align}
 & \Sigma_{\rm xc}^{\rm COHSEX} = \Sigma_{\rm xc}^{\rm SEX}+\Sigma_{\rm xc}^{\rm COH}, \nonumber \\
 & \Sigma_{\rm xc}^{\rm SEX}(\mathbf{x},\mathbf{x}') = -\sum_{k\sigma}^{\rm occ} \psi_{k\sigma}(\mathbf{x})\psi^*_{k\sigma}(\mathbf{x}')W(\mathbf{r},\mathbf{r}'), \nonumber \\
 & \Sigma_{\rm xc}^{\rm COH} (\mathbf{x},\mathbf{x}') = \frac{1}{2}\delta(\mathbf{x}, \mathbf{x}')[W(\mathbf{r},\mathbf{r}')-f_{\rm H}(\mathbf{r},\mathbf{r}')].
\end{align}
Like the HF self-energy, also the COHSEX one depends only on the density-matrix (when keeping $W$ constant, independent on the orbitals) 
and is then invariant under unitary rotations of the occupied manifold. Rewriting the $\delta$ function
using a completeness relation, $\delta(\mathbf{x},\mathbf{x}')=\sum_{k\sigma} \psi_{k\sigma}(\mathbf{x})\psi^*_{k\sigma}(\mathbf{x}')$,
and assuming again to work on a localized representation of the manifold to neglect off-diagonal 
contributions, the COHSEX matrix elements become  (see SI for a detailed derivation):
\begin{equation}
 \langle {i\sigma} | \Sigma^{\rm COHSEX}_{\rm xc} | {j\sigma'} \rangle \simeq \delta_{ij} \delta_{\sigma\sigma'}\left\{ \left(\frac{1}{2}-f_{i\sigma}\right) \langle n_{i\sigma} | W | n_{i\sigma} \rangle 
                    - \frac{1}{2} \langle n_{i\sigma} | f_{\rm H} | n_{i\sigma} \rangle  \right\} .
                    \label{eq:cohsex}
\end{equation}
The second term in the curly brackets is nothing but the self-Hartree of the orbital $i\sigma$, and the equation above 
matches exactly with Eq.~(\ref{eq:hkipz_r}). As already stated, so far we have only considered the Hartree and 
screening contributions to the potentials. On one side, this derivation highlights the role of the KC potentials as local and 
orbital dependent approximation to non-local self energies. On the other side, when a more advanced DFT functional is used as a starting 
point, also the xc potential and $f_{\rm xc}$ kernel play a role. 

The inclusion of $f_{\rm xc}$ transforms the test charge-test charge (RPA) dielectric function in the test charge-test electron response~\cite{ hybertsen_ab_1987,
del_sole_gwgamma_1994,onida_electronic_2002, bruneval_many-body_2005}. The electrons that participate to the screening have now an approximate
xc-hole surrounding them and the potential induced by the additional electron or hole includes xc-interactions 
and not only the classical Hartree term; 
the induced charge-density of the interacting system can be written~\cite{gross_local_1985,petersilka_excitation_1996} as
the charge density response of the auxiliary KS (non-interacting) system to an effective perturbation, i.e. $\delta \rho = \chi_0 \delta V_{\rm tot}$
where $\chi_0$ is the KS density-density response function. The effective potential $\delta V_{\rm tot}$ 
includes not only the external perturbation but also the self-consistent variation of the Hxc 
potential induced by the changes in the charge density, i.e. $\delta V_{\rm tot} = \delta V_{\rm ext} + \delta V_{\rm Hxc}$.
The inclusion of xc effects at the DFT level makes the link to a corresponding self-energy less trivial.
However, we argue that a screened KIPZ Hamiltonian built on a local or semi-local density functional may
be a good approximation to a more sophisticated self-energy operator where also simple (DFT based) 
vertex corrections are included (see discussion below).

Following Refs.~\citenum{hybertsen_electron_1986} and~\citenum{del_sole_gwgamma_1994} a zero-order approximation
for the self-energy can be defined using the DFT xc potential~\cite{casida_vxc_note}, i.e. $\Sigma_{\rm xc}(1,2)=\delta(1,2)V_{\rm xc}(1)$ 
[we used the compact notation 1=$(\mathbf{r}_1,t_1)$]; the first iteration of the Hedin's equation~\cite{hedin_new_1965}
leads to a GW$\Gamma$ approximation of the self energy that 
has an expression very similar to the RPA one, i.e. $\Sigma(1,2)=iG(1,2)\tilde{W}(1,2)$, but with 
a screened interaction that accounts for xc effects beyond the classical Hartree one: $\tilde{W}=\tilde{\varepsilon}^{-1}f_{\rm H}=f_{\rm H}[1-\chi_0f_{\rm Hxc}]^{-1}$.
Within this approximation Eq.~(\ref{eq:cohsex}) gets modified since now $W$ is replaced by $\tilde{W}$. The bare Coulomb interaction
is now screened by the test charge-test electron response function $\tilde{\varepsilon}^{-1}$, which is the one appearing also in the definition of the KC screening 
parameters, Eqs.~(\ref{eq:alpha_1}) and (\ref{eq:alpha_4}). Still a direct comparison of Eq.~(\ref{eq:cohsex}), including the improved screened interaction $\tilde{W}$, with
the full KIPZ matrix elements
\begin{equation}
 \langle {i\sigma} | v_{j\sigma', \rm xc}^{\rm rKIPZ(2)} | {j\sigma'} \rangle \simeq \delta_{ij}\delta_{\sigma\sigma'}\left\{  \langle \phi_{i\sigma} | v^{\rm DFT}_{\sigma,\rm xc} | \phi_{i\sigma} \rangle  + 
 \left(\frac{1}{2}-f_{i\sigma}\right) \langle n_{i\sigma} | \mathcal{F}^{\sigma\sigma}_{\rm Hxc} | n_{i\sigma} \rangle - E_{\rm Hxc}[n_{i\sigma}]\right\}
\end{equation}
is not trivial, and is hindered by the presence of the xc energy and potential of the underlying DFT functional.
Nevertheless, the similarity between the two approaches is apparent also in this case, and highlights 
the physical ingredients present in the KIPZ orbital-dependent potentials.

\section{Computational Details}
In the next Sections we provide numerical results from KC calculations on the GW100 test set.
This set has been introduced~\cite{van_setten_gw100_2015} to benchmark and validate different GW implementations
and is made of 100 small closed-shell molecules with diverse chemical bonding. 
For consistency we adopt the same molecular geometries provided in the original work.
In our analysis we focus on the first ionization potential (IP) for which accurate CCSD(T) results
are available; for a subset of molecules we also look at transitions from deeper valence states 
for which a comparison with experimental data (photoemission spectroscopy) is available. 

Standard DFT calculations presented here have been performed using the PWSCF code in the 
{\sc Quantum ESPRESSO}~\cite{giannozzi_quantum_2009,giannozzi_advanced_2017} distribution. 
Extensive convergence tests on the energy cut-off for the 
plane-wave expansion, as well on the size of the supercell have been performed to 
ensure a converged value of the first IP within 10 meV. The supercell size convergence has
been facilitated thanks to the use of reciprocal-space counter-charge corrections.~\cite{li_electronic_2011,martyna_reciprocal_1999}
The electron-ion interactions have been modeled using Optimized Norm-Conserving Vanderbilt (ONCV) 
pseudopotentials~\cite{hamann_optimized_2013} as developed by Schlipf and Gygi~\cite{ schlipf_optimization_2015, ONCV_website}.
The KC calculations have been performed using a modified version of the {\sc Quantum ESPRESSO}
CP code and adopt the Perdew, Burke and Ernzerhof (PBE) exchange-correlation functional~\cite{perdew_generalized_1996}
as the underlying base functional. The minimization of the orbital-density dependent 
functionals has been done on the space of complex-valued wavefunctions; the strategy 
we use for the optimization follows the ensemble-DFT algorithm~\cite{marzari_ensemble_1997}
and consist of two nested loops~\cite{borghi_variational_2015}; in the inner loop
we search for the optimal unitary transformation that minimizes the orbital-density
dependent part of the energy functional at fixed manifold; in the outer loop
a standard conjugate-gradients algorithm is used to optimize the manifold.
The orbital-density dependent screening coefficients have been computed according to 
Eq.~\eqref{eq:alpha_1} using the linear-response scheme described in Ref.~\citenum{colonna_screening_2018}; 
for the case of KIPZ, where this scheme is not applicable due the lack of an implementation of 
linear-response analytical derivatives for the PZ-SIC functional, we resort to the finite difference approach outlined 
in Ref.~\citenum{nguyen_koopmans-compliant_2017}.   

The theoretical IPs are defined as minus the eigenvalue of the corresponding molecular orbital~\cite{ip_ehomo_note} 
for all the theoretical methods reported, except for 
$\Delta$SCF. In the latter case only the first IP has been computed, defined as 
the energy difference between the neutral molecule and its cation.

\section{Results}

\begin{figure}
\begin{center}
\includegraphics[angle=-90,width=0.9\textwidth]{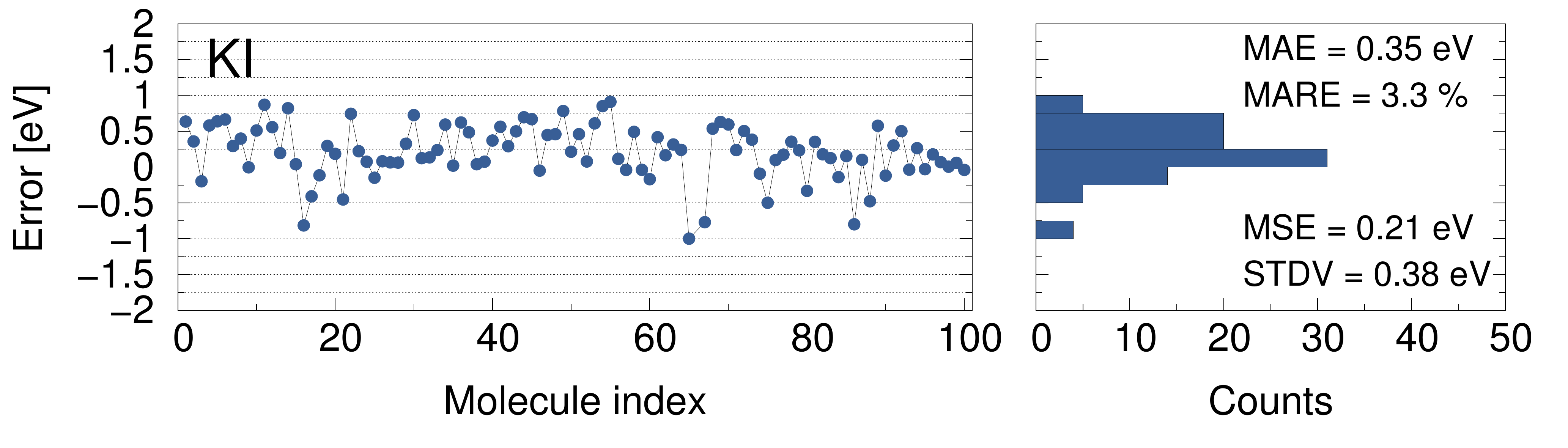}
\includegraphics[angle=-90,width=0.9\textwidth]{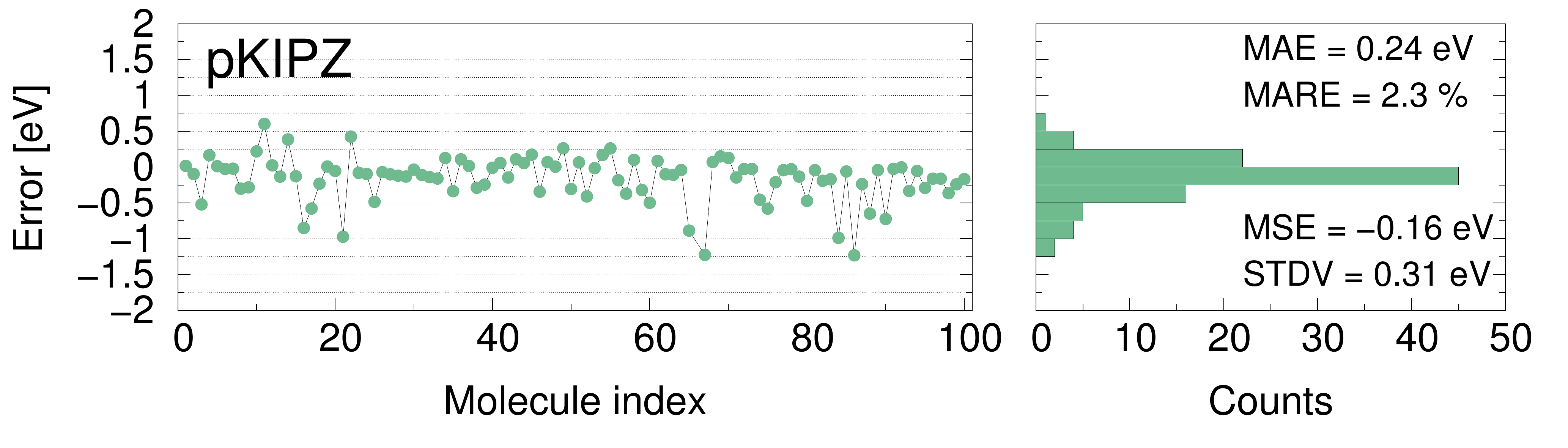}
\includegraphics[angle=-90,width=0.9\textwidth]{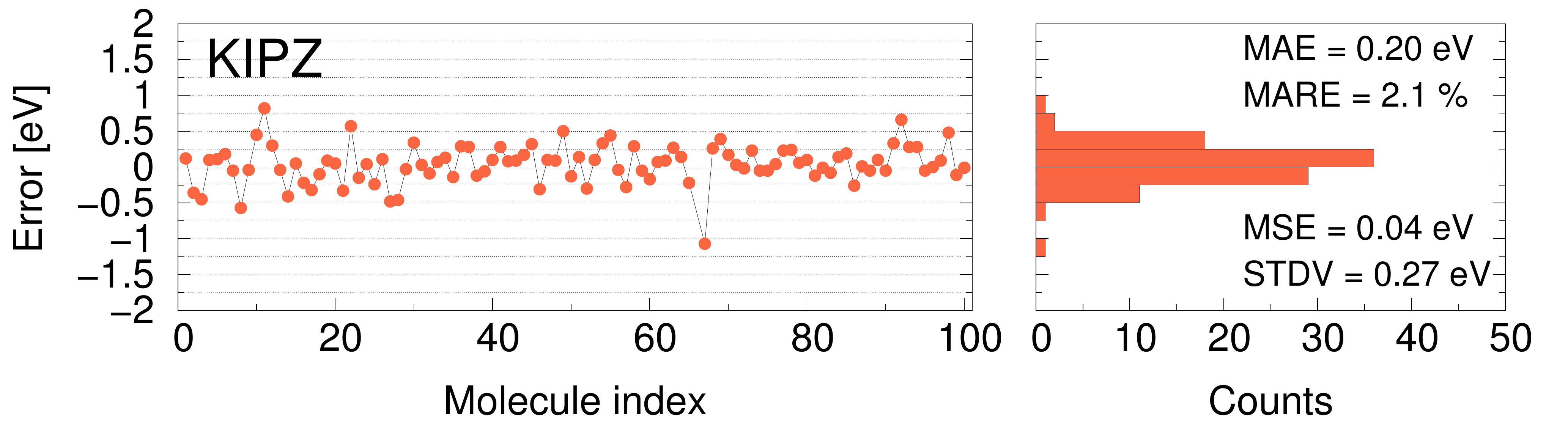}
\caption{Errors (left panels) on the first ionization potentials with respect to CCSD(T) $\Delta$SCF results~\cite{krause_coupled-cluster_2015}
and their distributions (right panel) at different levels of the theory.
Top panel KI, central panel perturbative KIPZ and bottom panel full KIPZ (see text for details).}
\label{fig.histo_err}
\end{center}
\end{figure}

\subsection{First ionization potential}
In this Section we analyze the average performance of the KC functionals 
described in Sec.~\ref{sec:KC_func}. To begin with, we focus on the first ionization potential (IP),
for which accurate quantum-chemistry calculations (such as, e.g., $\Delta$SCF
CCSD(T) at the def2-TZVPP level) are available in the literature.~\cite{krause_coupled-cluster_2015}
Comparison against high-accuracy methods allows one to remove the experimental
uncertainties and strictly focus on the failures of the theory. To quantify the 
agreement with respect to the CCSD(T) reference we look at the Mean Absolute Error (MAE), 
the Mean Absolute Relative Error (MARE), the Mean Signed Error (MSE), and the Standard Deviation (STDV), defined as 
\begin{align}
& {\rm MAE}   = \frac{1}{N}\sum_{i=1}^N |\Delta{\rm IP}_i| \nonumber \\
& {\rm MARE}  = \frac{1}{N}\sum_{i=1}^N |\Delta{\rm IP}_i/{\rm IP}_i^{\rm R}| \nonumber \\
& {\rm MSE}   = \frac{1}{N}\sum_{i=1}^N \Delta{\rm IP}_i \nonumber \\
& {\rm STDV}  = \sqrt{\frac{1}{N}\sum_{i=1}^N (\Delta{\rm IP}_i -{\rm MSE})^2} 
\end{align}
with $\Delta{\rm IP} = {\rm IP} - {\rm IP^{\rm R}}$ the error in the ionization
potential ${\rm IP}$ with respect to the reference [CCSD(T)] value ${\rm IP}^{\rm R}$.

In Fig.~\ref{fig.histo_err} we present the results for the first ionization potentials calculated 
within the KI and KIPZ functionals (top and bottom panels, respectively). We also report results
for perturbative KIPZ calculations (pKIPZ in the plots and tables, middle panel) 
where the KIPZ correction is computed on top of the KI minimizing orbitals, thus 
neglecting self-consistency effects at the KIPZ level. 
We recall here that the KI total energy at integer occupation numbers coincides with that of the underlying density functional 
and thus KI preserves its unitary invariance under rotation of the manifold. As already discussed 
in previous works~\cite{borghi_koopmans-compliant_2014,colonna_screening_2018} we choose 
a specific manifold by introducing an infinitesimally small PZ-SIC
contribution to the KI energy. This allows us to (1) unambiguously define the manifold, since the small PZ-SIC term
breaks the unitary invariance, and to (2) localize the orbitals without modifying the ground state energy.

The performance of these three KC functionals is summarized in the right panels of Fig.~\ref{fig.histo_err}. Our
calculations find that KI slightly overestimates the ionization potential by 0.21 eV, on average (MSE). 
Simply applying the KIPZ correction in a perturbative way significantly improves the results reducing the 
MAE by 0.11 eV and narrowing the distribution of the error. A MSE of -0.16 eV indicates an underestimation 
of the IP on average, with an opposite trend with respect to the KI calculation. 
A full KIPZ calculation reduces further both the MAE (0.20 eV) and the STDV (0.27 eV), 
 the error distribution being nicely centered around zero with a MSE of 0.04 eV. 
Notwithstanding the further improvement obtained within KIPZ, we note that the largest change happens when 
passing from KI to pKIPZ. This seems to suggest that the main reason for the improvement
is in the KIPZ Hamiltonian and only secondarily in the self-consistency effects (changes in the orbitals and orbital densities).
Indeed, the definition of the KI functional can be thought of as the limit of the full 
KIPZ functional when the PZ-SIC contribution is sent to zero~\cite{borghi_koopmans-compliant_2014} (see Eq.~\ref{eq:uPi_KIPZ}),
and thus further justifies the use of the KI manifold as the starting point for the perturbative KIPZ calculation. 

\begin{figure*}
\begin{minipage}{\textwidth} 
\includegraphics[width=\textwidth]{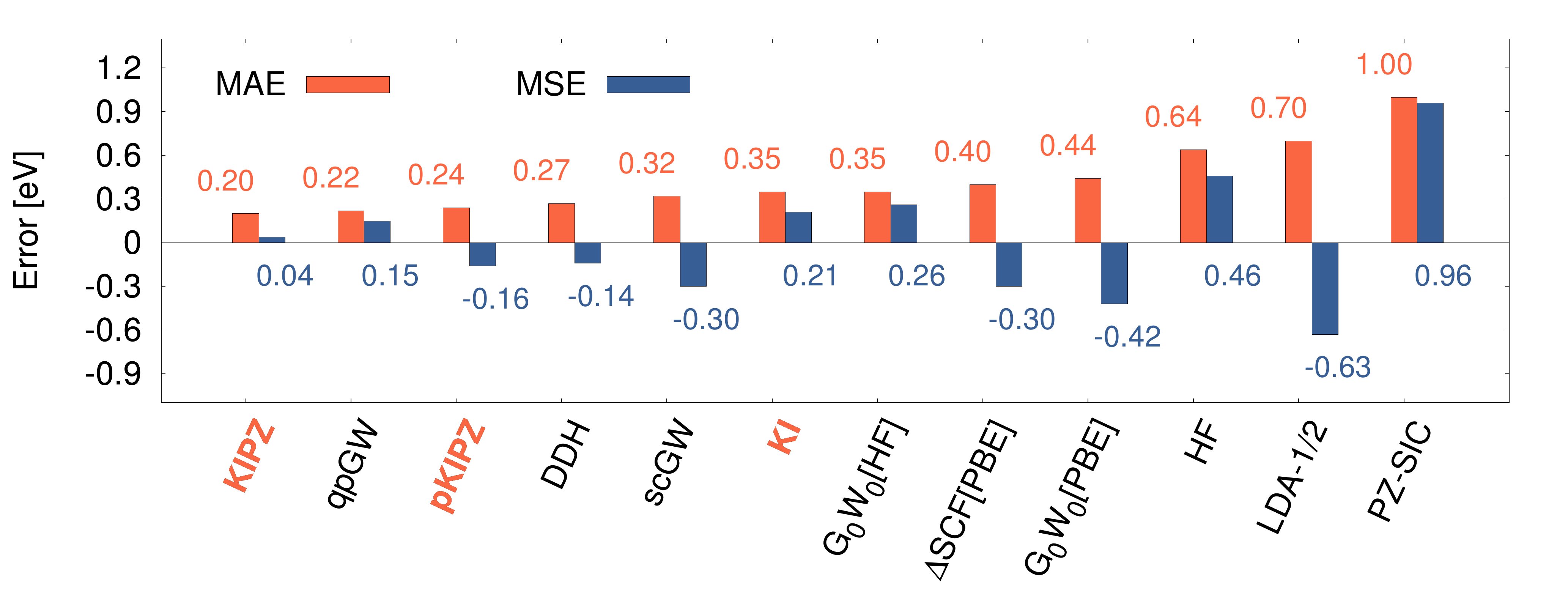}
 \captionof{figure}{Comparison of the average performance of the KC functionals with other electronic structure methods.
  HF and LDA-1/2 results from Ref.~\citenum{pela_lda-1/2_2018}, DDH from Ref.~\citenum{brawand_performance_2017}, 
  G$_0$W$_0$[PBE] from Ref.~\citenum{govoni_gw100_2018} and G$_0$W$_0$[HF], scGW and qpGW from Ref.~\citenum{caruso_benchmark_2016}.
  Red labels on the horizontal axis are used to highlight KC functionals.}
\label{fig:comparison}
\end{minipage}
\end{figure*}

In Fig.~\ref{fig:comparison} we compare KC functionals with other electronic structure methods characterized by 
different level of complexity: the LDA-1/2 approach~\cite{ferreira_approximation_2008}, a purely KS 
scheme based on a local potential; Hartree-Fock (HF) and dielectric dependent hybrid (DDH)
functionals~\cite{skone_self-consistent_2014, skone_nonempirical_2016, brawand_generalization_2016},
 accounting for the non-locality of the potential; and various GW schemes 
ranging from perturbative G$_0$W$_0$ to fully self-consistent GW, taking also into account  the (dynamical)
screening in the effective single particle potential (self-energy). In terms of complexity, KC functionals, 
with their local but orbital-dependent potentials, stand in between LDA-1/2 and the hybrid
functionals. It is remarkable then that the accuracy they achieve is comparable or actually superior to that 
of more sophisticated electronic structure approaches with the smallest error compared to CCSD(T). 

It is fair to say that although the KC potentials are local, the computation of all the 
ingredients needed, and in particular of the screening coefficients, requires additional effort. 
This is apparent from Eq.~\eqref{eq:alpha_1}, where screening is defined as an orbital-dependent
average of the static DFT dielectric matrix. This resonates with DDH functionals, where the mixing parameter is 
obtained from a screened-exchange calculation and thus ultimately requires the knowledge of the (static)
RPA dielectric matrix. However, it is important to stress that the scheme we use to compute the $\{\alpha_{i\sigma}\}$~\cite{colonna_screening_2018}
does not requires to build up the whole dielectric matrix [neither in a PW basis (as done in standard implementations) nor in any other 
optimal basis (as done for DDH~\cite{brawand_generalization_2016,brawand_performance_2017})]; we instead compute ``on the fly'' the action of the dielectric matrix on
the orbital density that we are considering. The evaluation of each screening coefficient requires
thus a linear-response calculation that scales roughly as an electronic self-consistent DFT calculation.
Moreover, since the variational orbitals are Wannier-like, they form groups
with very similar screening coefficients; although not exploited here, this feature would allow one 
to greatly reduce the number of independent linear response calculations~\cite{colonna_screening_2018, nguyen_koopmans-compliant_2017} 
and speed up and/or streamline the calculations.

\subsection{Deeper valence states}\label{sec:low_lying_states}

\begin{figure}[ht]
    \centering
    \hspace*{1cm}\includegraphics[width=0.45\textwidth]{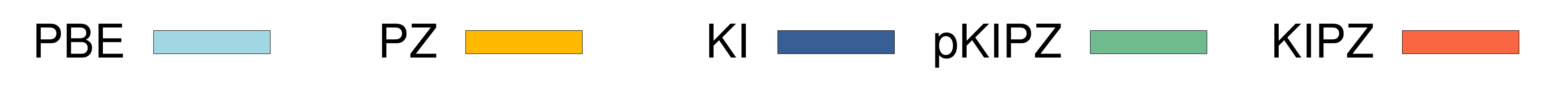}
    \includegraphics[width=0.5\textwidth]{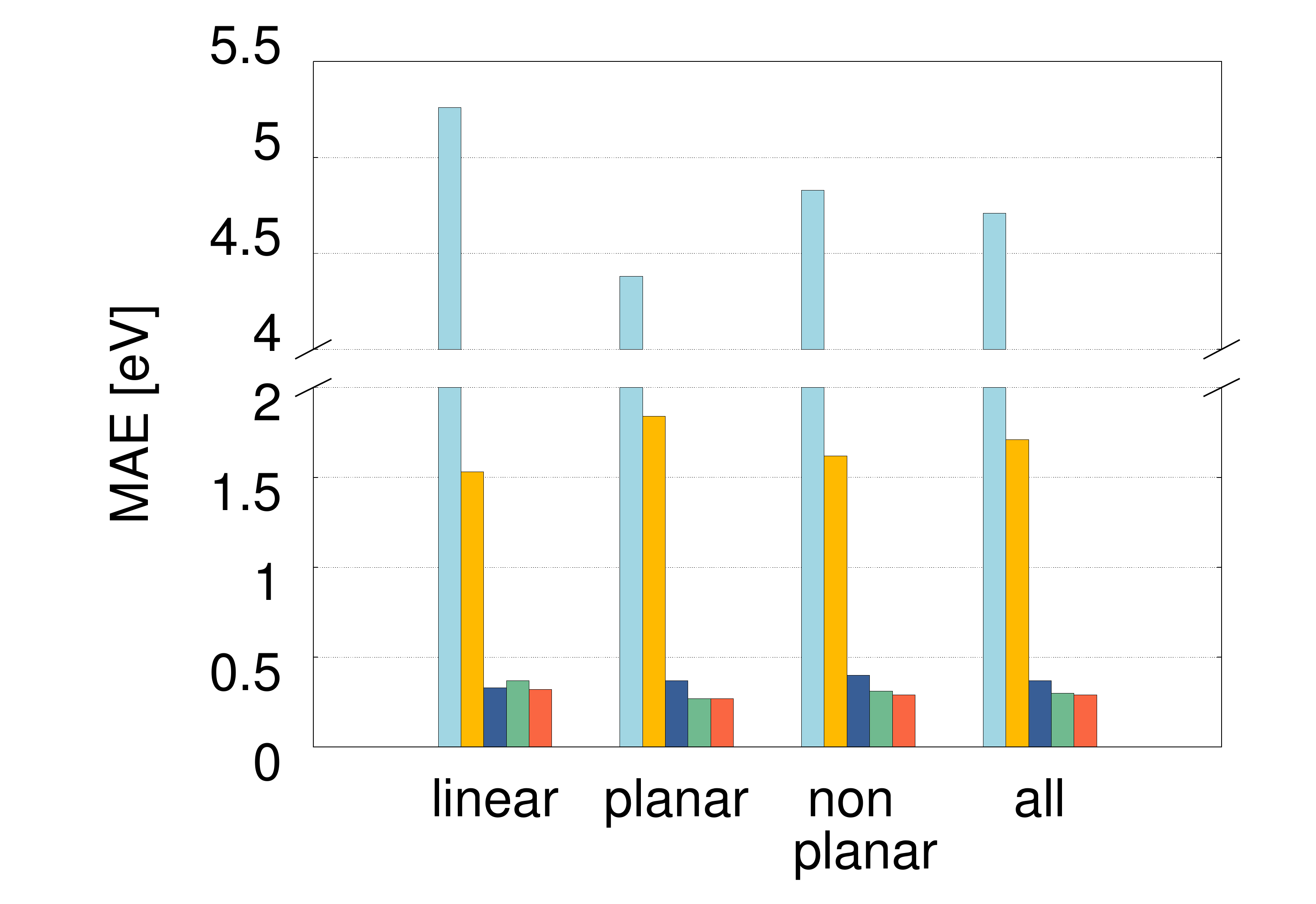}
     \caption{MAE on the deeper valence state IPs with respect to experimental values for different functionals.
      The average is first performed separately on 18 IPs from a subset of linear molecules (N$_2$, F$_2$, HF, CO, HCN, C$_2$H$_2$), on 38 IPs
      form a subset of planar molecules (H$_2$O, C$_2$H$_4$, O$_3$, HCOOH, benzene and pyridine) and on 21 IPs from 
      a subset of non-planar molecules (NH$_3$, PH$_3$, C$_2$H$_6$, CF$_4$, CCl$_4$), and then on the 77 transitions all together.
      The complete list of vertical ionization potentials considered is given in SI.}
    \label{fig:lowerstates_mae}
\end{figure}

\begin{table}[t]
\caption{MAE and MSE from experiments on the deeper valence state IPs from the subset of 18 molecules
considered in Sec.~\ref{sec:low_lying_states} and for different functionals. The average is first performed
separately on the 18 IPs from the linear molecules, on the 38 IPs from the planar molecules, and on the 21
IPs from the non-planar molecules, and then on all the 77 transitions. The MAE and MSE only on the 18 first IPs 
is also given in the last two lines.}
\label{tab:lowerstates_mae}
 \begin{tabular}{l C{1.2cm} C{1.5cm} C{1.6cm}  C{1.5cm}  C{1.5cm}  C{1.5cm}  } 
\hline
\hline 
\T\B 
Type of Molecule &  	 	& PBE   	& PZ-SIC 	& KI		& pKIPZ		& KIPZ	\\
\hline
\T 
Linear   	   &  MAE	&    5.26 	&  1.53 	&  0.33 	&   0.37	&  0.32  \\
		   &  MSE 	&   -5.26	&  1.47 	&  0.08 	&  -0.21	&  0.06  \\
		   \T
Planar   	   &  MAE	&    4.38	&  1.84 	&  0.37 	&   0.27	&  0.27  \\
		   &  MSE 	&   -4.38	&  1.84 	&  0.22 	&  -0.06	&  0.06  \\
		   \T
Non-planar         &  MAE	&    4.83	&  1.62 	&  0.40 	&   0.31 	&  0.29  \\
		   &  MSE 	&   -4.83	&  1.62 	&  0.11 	&  -0.08	&  0.08  \\
		   \hline
	 	   \T
All IPs (77)    	&  MAE	&    4.71 	&  1.71 	&  0.37 	&   0.30	&  0.29  \\
			\B
			&  MSE 	&   -4.71	&  1.69 	&  0.16 	&  -0.10	&  0.06  \\
			\T
First IPs (18)   	&  MAE	&    4.62 	&  1.21 	&  0.37 	&   0.32	&  0.23  \\
			\B
			&  MSE 	&   -4.62	&  1.20 	&  0.22 	&  -0.17	&  0.06  \\		   
			\hline\hline
\end{tabular}
\end{table}

In this Section we report results concerning the use of Koopmans functionals to evaluate low-lying single particle energies. The linearity 
condition typical of all the KC flavors applies to the entire electronic manifold (and not just to the highest occupied 
orbital, as in exact DFT), and one thus expects meaningful corrections also for states different from the HOMO.
We considered 77 vertical ionization potentials from a subset of the molecules studied, for which experimental 
results are available in Refs.~\citenum{chong_interpretation_2002} and \citenum{shapley_pw86pw91_2001} (and references
therein).
We group the molecules according to their chemical and structural properties: the first group comprises 
6 linear molecules including simple dimers (N$_2$ and F$_2$), polar molecules (HF and CO), and two molecules with a
strong bond, i.e. hydrogen cyanide (C-N triple bond) and the simplest alkyne C$_2$H$_2$ (C-C triple bond).
The second group is made by 6 planar molecules (O$_3$, H$_2$O, HCOOH, C$_2$H$_4$, benzene and pyridine)
of increasing size, with benzene and pyridine representative for aromatic, organic molecules. 
The last group comprises 6 small non-planar molecules (NH$_3$, PH$_3$, C$_2$H$_6$, CF$_4$, CCl$_4$). 
We mainly consider transitions with binding energies smaller than 21.2 eV, i.e. accessible via He(I) ultraviolet 
photoemission spectroscopy, with few exceptions where higher binding-energy transitions are also included (see
SI for the complete list of vertical ionization potentials considered).

In Fig.~\ref{fig:lowerstates_mae} and Tab.~\ref{tab:lowerstates_mae} we show the average performance of PBE, PZ-SIC and the KC 
functionals with respect to experimental results. For the IPs considered we observe the same trend found for the case
of the HOMO when moving from standard density functionals to PZ-SIC and eventually to KC; the PBE eigenvalues greatly underestimate
the experimental values while PZ-SIC tends to overestimate transition energies. 
All the KC flavors show a more balanced performance with pKIPZ and KIPZ having very similar MAE 
and MSE. Most importantly, we do not observe any significant difference in the average performance when deeper states are 
concerned. This is clear from the comparison of the last two lines in Tab.~\ref{tab:lowerstates_mae} where the average 
over all transitions and the average over only the first ionization energies are listed. 
Only for PZ-SIC an increase of the errors by $\sim$ 0.5 eV is observed when deeper states are included in the average; 
for all the KC functionals very small changes are observed (0.00 eV, -0.02 eV, +0.06 eV in the MAE for KI,
pKIPZ and KIPZ, respectively), highlighting the predictive capabilities of Koopmans-compliant functional 
also for states different from the HOMO.

\section{Conclusions}
In conclusion, we have investigated in detail Koopmans-compliant functionals highlighting similarity 
and differences with respect to other well established electronic structure methods, and tested the
full GW100 protocol. 
We show that the local and orbital-density dependent potentials typical of  
KC functionals can be thought of as a simplified yet accurate approximation to the 
self-energy operator beyond the GW approximation. We stressed once again the importance of the local nature 
of the variational orbitals that ultimately is responsible for the possibility to 
map the non-local self-energy operator into local but orbital dependent 
potentials. We ascribe the remarkably good performance of KC functionals (and in 
particular of KIPZ) to the inclusion of local vertex corrections which allow for
an improved description of the screening processes and self-energies 
beyond the RPA approximation, typical of standard GW approaches. 
This theoretical analysis is supported by the numerical calculations of ionization potentials 
for the large (and standardized) GW100 set of molecules. The KIPZ functionals show remarkably good performance 
with a MAE or 0.20 eV, significantly better than all standard perturbative G$_0$W$_0$ 
results and in line with GW calculations when some sort of self-consistency is considered. 
On the basis of the theoretical analysis presented here we argue that the inclusion of 
vertex corrections may be important to further close the gap between Green's function methods results and CCCSD(T).
In the KC functionals this is done at the DFT level with a local vertex function. 
When deeper valence states are concerned, no significant change in the average performance 
of the KC functionals is observed, highlighting the effectiveness of the orbital-density 
dependent potentials in accurately describing the whole electronic structure manifold and
not just the first ionization potential.

\begin{acknowledgement}
This research was supported by the Swiss National Science
Foundation (SNSF), through Grant No. 200021-179139, and
its National Centre of Competence in Research (NCCR)
MARVEL. We acknowledge the computing center
of EPFL for high performance computing resources. 
\end{acknowledgement}

\begin{suppinfo}\label{Supplemental_material}
The Supporting Information contains  details on the derivation of Eq.~\ref{eq:cohsex} in Sec.~\ref{sec:cohsex};
the complete list of the first ionization potentials for the entire set of 100 molecule (Tab.~SI-1) and the list
of 77 deeper vertical ionization potentials (Tabs.~SI-2, SI-3 and SI-4) for the subset of molecules considered in Sec~\ref{sec:low_lying_states}.
\end{suppinfo}


\begin{mcitethebibliography}{77}
\providecommand*\natexlab[1]{#1}
\providecommand*\mciteSetBstSublistMode[1]{}
\providecommand*\mciteSetBstMaxWidthForm[2]{}
\providecommand*\mciteBstWouldAddEndPuncttrue
  {\def\EndOfBibitem{\unskip.}}
\providecommand*\mciteBstWouldAddEndPunctfalse
  {\let\EndOfBibitem\relax}
\providecommand*\mciteSetBstMidEndSepPunct[3]{}
\providecommand*\mciteSetBstSublistLabelBeginEnd[3]{}
\providecommand*\EndOfBibitem{}
\mciteSetBstSublistMode{f}
\mciteSetBstMaxWidthForm{subitem}{(\alph{mcitesubitemcount})}
\mciteSetBstSublistLabelBeginEnd
  {\mcitemaxwidthsubitemform\space}
  {\relax}
  {\relax}

\bibitem[Yu and Zunger(2012)Yu, and Zunger]{yu_identification_2012}
Yu,~L.; Zunger,~A. Identification of {Potential} {Photovoltaic} {Absorbers}
  {Based} on {First}-{Principles} {Spectroscopic} {Screening} of {Materials}.
  \emph{Phys. Rev. Lett.} \textbf{2012}, \emph{108}, 068701\relax
\mciteBstWouldAddEndPuncttrue
\mciteSetBstMidEndSepPunct{\mcitedefaultmidpunct}
{\mcitedefaultendpunct}{\mcitedefaultseppunct}\relax
\EndOfBibitem
\bibitem[E.~Castelli \latin{et~al.}(2012)E.~Castelli, Olsen, Datta, D.~Landis,
  Dahl, S.~Thygesen, and W.~Jacobsen]{castelli_computational_2012}
E.~Castelli,~I.; Olsen,~T.; Datta,~S.; D.~Landis,~D.; Dahl,~S.;
  S.~Thygesen,~K.; W.~Jacobsen,~K. Computational screening of perovskite metal
  oxides for optimal solar light capture. \emph{En. Environ. Sci.}
  \textbf{2012}, \emph{5}, 5814--5819\relax
\mciteBstWouldAddEndPuncttrue
\mciteSetBstMidEndSepPunct{\mcitedefaultmidpunct}
{\mcitedefaultendpunct}{\mcitedefaultseppunct}\relax
\EndOfBibitem
\bibitem[Jain \latin{et~al.}(2016)Jain, Shin, and
  Persson]{jain_computational_2016}
Jain,~A.; Shin,~Y.; Persson,~K.~A. Computational predictions of energy
  materials using density functional theory. \emph{Nature Reviews Materials}
  \textbf{2016}, \emph{1}, 15004\relax
\mciteBstWouldAddEndPuncttrue
\mciteSetBstMidEndSepPunct{\mcitedefaultmidpunct}
{\mcitedefaultendpunct}{\mcitedefaultseppunct}\relax
\EndOfBibitem
\bibitem[Pham \latin{et~al.}(2017)Pham, Govoni, Seidel, Bradforth, Schwegler,
  and Galli]{pham_electronic_2017}
Pham,~T.~A.; Govoni,~M.; Seidel,~R.; Bradforth,~S.~E.; Schwegler,~E.; Galli,~G.
  Electronic structure of aqueous solutions: {Bridging} the gap between theory
  and experiments. \emph{Sci. Adv.} \textbf{2017}, \emph{3}, e1603210\relax
\mciteBstWouldAddEndPuncttrue
\mciteSetBstMidEndSepPunct{\mcitedefaultmidpunct}
{\mcitedefaultendpunct}{\mcitedefaultseppunct}\relax
\EndOfBibitem
\bibitem[Onida \latin{et~al.}(2002)Onida, Reining, and
  Rubio]{onida_electronic_2002}
Onida,~G.; Reining,~L.; Rubio,~A. Electronic excitations: density-functional
  versus many-body {Green}'s-function approaches. \emph{Rev. Mod. Phys.}
  \textbf{2002}, \emph{74}, 601--659\relax
\mciteBstWouldAddEndPuncttrue
\mciteSetBstMidEndSepPunct{\mcitedefaultmidpunct}
{\mcitedefaultendpunct}{\mcitedefaultseppunct}\relax
\EndOfBibitem
\bibitem[Perdew \latin{et~al.}(1982)Perdew, Parr, Levy, and
  Balduz]{perdew_density-functional_1982}
Perdew,~J.~P.; Parr,~R.~G.; Levy,~M.; Balduz,~J.~L. Density-{Functional}
  {Theory} for {Fractional} {Particle} {Number}: {Derivative} {Discontinuities}
  of the {Energy}. \emph{Phys. Rev. Lett.} \textbf{1982}, \emph{49},
  1691--1694\relax
\mciteBstWouldAddEndPuncttrue
\mciteSetBstMidEndSepPunct{\mcitedefaultmidpunct}
{\mcitedefaultendpunct}{\mcitedefaultseppunct}\relax
\EndOfBibitem
\bibitem[Almbladh and von Barth(1985)Almbladh, and von
  Barth]{almbladh_exact_1985}
Almbladh,~C.-O.; von Barth,~U. Exact results for the charge and spin densities,
  exchange-correlation potentials, and density-functional eigenvalues.
  \emph{Phys. Rev. B} \textbf{1985}, \emph{31}, 3231--3244\relax
\mciteBstWouldAddEndPuncttrue
\mciteSetBstMidEndSepPunct{\mcitedefaultmidpunct}
{\mcitedefaultendpunct}{\mcitedefaultseppunct}\relax
\EndOfBibitem
\bibitem[Chong \latin{et~al.}(2002)Chong, Gritsenko, and
  Baerends]{chong_interpretation_2002}
Chong,~D.~P.; Gritsenko,~O.~V.; Baerends,~E.~J. Interpretation of the
  {Kohn}{\textendash}{Sham} orbital energies as approximate vertical ionization
  potentials. \emph{J. Chem. Phys.} \textbf{2002}, \emph{116}, 1760--1772\relax
\mciteBstWouldAddEndPuncttrue
\mciteSetBstMidEndSepPunct{\mcitedefaultmidpunct}
{\mcitedefaultendpunct}{\mcitedefaultseppunct}\relax
\EndOfBibitem
\bibitem[Cococcioni and de~Gironcoli(2005)Cococcioni, and
  de~Gironcoli]{cococcioni_linear_2005}
Cococcioni,~M.; de~Gironcoli,~S. Linear response approach to the calculation of
  the effective interaction parameters in the
  \${\textbackslash}mathrm\{{LDA}\}+{\textbackslash}mathrm\{{U}\}\$ method.
  \emph{Phys. Rev. B} \textbf{2005}, \emph{71}, 035105\relax
\mciteBstWouldAddEndPuncttrue
\mciteSetBstMidEndSepPunct{\mcitedefaultmidpunct}
{\mcitedefaultendpunct}{\mcitedefaultseppunct}\relax
\EndOfBibitem
\bibitem[Kulik \latin{et~al.}(2006)Kulik, Cococcioni, Scherlis, and
  Marzari]{kulik_density_2006}
Kulik,~H.~J.; Cococcioni,~M.; Scherlis,~D.~A.; Marzari,~N. Density {Functional}
  {Theory} in {Transition}-{Metal} {Chemistry}: {A} {Self}-{Consistent}
  {Hubbard} \${U}\$ {Approach}. \emph{Phys. Rev. Lett.} \textbf{2006},
  \emph{97}, 103001\relax
\mciteBstWouldAddEndPuncttrue
\mciteSetBstMidEndSepPunct{\mcitedefaultmidpunct}
{\mcitedefaultendpunct}{\mcitedefaultseppunct}\relax
\EndOfBibitem
\bibitem[Dabo \latin{et~al.}(2009)Dabo, Cococcioni, and
  Marzari]{dabo_non-koopmans_2009}
Dabo,~I.; Cococcioni,~M.; Marzari,~N. Non-{Koopmans} {Corrections} in
  {Density}-functional {Theory}: {Self}-interaction {Revisited}.
  \emph{arXiv:0901.2637 [cond-mat]} \textbf{2009}, arXiv: 0901.2637\relax
\mciteBstWouldAddEndPuncttrue
\mciteSetBstMidEndSepPunct{\mcitedefaultmidpunct}
{\mcitedefaultendpunct}{\mcitedefaultseppunct}\relax
\EndOfBibitem
\bibitem[Dabo \latin{et~al.}(2010)Dabo, Ferretti, Poilvert, Li, Marzari, and
  Cococcioni]{dabo_koopmans_2010}
Dabo,~I.; Ferretti,~A.; Poilvert,~N.; Li,~Y.; Marzari,~N.; Cococcioni,~M.
  Koopmans' condition for density-functional theory. \emph{Phys. Rev. B}
  \textbf{2010}, \emph{82}, 115121\relax
\mciteBstWouldAddEndPuncttrue
\mciteSetBstMidEndSepPunct{\mcitedefaultmidpunct}
{\mcitedefaultendpunct}{\mcitedefaultseppunct}\relax
\EndOfBibitem
\bibitem[Stein \latin{et~al.}(2010)Stein, Eisenberg, Kronik, and
  Baer]{stein_fundamental_2010}
Stein,~T.; Eisenberg,~H.; Kronik,~L.; Baer,~R. Fundamental Gaps in Finite
  Systems from Eigenvalues of a Generalized Kohn-Sham Method. \emph{Phys. Rev.
  Lett.} \textbf{2010}, \emph{105}, 266802\relax
\mciteBstWouldAddEndPuncttrue
\mciteSetBstMidEndSepPunct{\mcitedefaultmidpunct}
{\mcitedefaultendpunct}{\mcitedefaultseppunct}\relax
\EndOfBibitem
\bibitem[Zheng \latin{et~al.}(2011)Zheng, Cohen, Mori-S\'anchez, Hu, and
  Yang]{zheng_improving_2011}
Zheng,~X.; Cohen,~A.~J.; Mori-S\'anchez,~P.; Hu,~X.; Yang,~W. Improving Band
  Gap Prediction in Density Functional Theory from Molecules to Solids.
  \emph{Phys. Rev. Lett.} \textbf{2011}, \emph{107}, 026403\relax
\mciteBstWouldAddEndPuncttrue
\mciteSetBstMidEndSepPunct{\mcitedefaultmidpunct}
{\mcitedefaultendpunct}{\mcitedefaultseppunct}\relax
\EndOfBibitem
\bibitem[Kraisler and Kronik(2013)Kraisler, and
  Kronik]{kraisler_piecewise_2013}
Kraisler,~E.; Kronik,~L. Piecewise {Linearity} of {Approximate} {Density}
  {Functionals} {Revisited}: {Implications} for {Frontier} {Orbital}
  {Energies}. \emph{Phys. Rev. Lett.} \textbf{2013}, \emph{110}, 126403\relax
\mciteBstWouldAddEndPuncttrue
\mciteSetBstMidEndSepPunct{\mcitedefaultmidpunct}
{\mcitedefaultendpunct}{\mcitedefaultseppunct}\relax
\EndOfBibitem
\bibitem[G{\"o}rling(2015)]{gorling_exchange-correlation_2015}
G{\"o}rling,~A. Exchange-correlation potentials with proper discontinuities for
  physically meaningful {Kohn}-{Sham} eigenvalues and band structures.
  \emph{Phys. Rev. B} \textbf{2015}, \emph{91}, 245120\relax
\mciteBstWouldAddEndPuncttrue
\mciteSetBstMidEndSepPunct{\mcitedefaultmidpunct}
{\mcitedefaultendpunct}{\mcitedefaultseppunct}\relax
\EndOfBibitem
\bibitem[Li \latin{et~al.}(2015)Li, Zheng, Cohen, Mori-S\'anchez, and
  Yang]{li_local_2015}
Li,~C.; Zheng,~X.; Cohen,~A.~J.; Mori-S\'anchez,~P.; Yang,~W. Local Scaling
  Correction for Reducing Delocalization Error in Density Functional
  Approximations. \emph{Phys. Rev. Lett.} \textbf{2015}, \emph{114},
  053001\relax
\mciteBstWouldAddEndPuncttrue
\mciteSetBstMidEndSepPunct{\mcitedefaultmidpunct}
{\mcitedefaultendpunct}{\mcitedefaultseppunct}\relax
\EndOfBibitem
\bibitem[Kronik \latin{et~al.}(2012)Kronik, Stein, Refaely-Abramson, and
  Baer]{kronik_excitation_2012}
Kronik,~L.; Stein,~T.; Refaely-Abramson,~S.; Baer,~R. Excitation Gaps of
  Finite-Sized Systems from Optimally Tuned Range-Separated Hybrid Functionals.
  \emph{J. Chem. Theory Comput.} \textbf{2012}, \emph{8}, 1515--1531\relax
\mciteBstWouldAddEndPuncttrue
\mciteSetBstMidEndSepPunct{\mcitedefaultmidpunct}
{\mcitedefaultendpunct}{\mcitedefaultseppunct}\relax
\EndOfBibitem
\bibitem[Refaely-Abramson \latin{et~al.}(2012)Refaely-Abramson, Sharifzadeh,
  Govind, Autschbach, Neaton, Baer, and
  Kronik]{refaely-abramson_quasiparticle_2012}
Refaely-Abramson,~S.; Sharifzadeh,~S.; Govind,~N.; Autschbach,~J.;
  Neaton,~J.~B.; Baer,~R.; Kronik,~L. Quasiparticle Spectra from a Nonempirical
  Optimally Tuned Range-Separated Hybrid Density Functional. \emph{Phys. Rev.
  Lett.} \textbf{2012}, \emph{109}, 226405\relax
\mciteBstWouldAddEndPuncttrue
\mciteSetBstMidEndSepPunct{\mcitedefaultmidpunct}
{\mcitedefaultendpunct}{\mcitedefaultseppunct}\relax
\EndOfBibitem
\bibitem[Li \latin{et~al.}(2018)Li, Zheng, Su, and Yang]{li_localized_2018}
Li,~C.; Zheng,~X.; Su,~N.~Q.; Yang,~W. Localized orbital scaling correction for
  systematic elimination of delocalization error in density functional
  approximations. \emph{Natl. Sci. Rev.} \textbf{2018}, \relax
\mciteBstWouldAddEndPunctfalse
\mciteSetBstMidEndSepPunct{\mcitedefaultmidpunct}
{}{\mcitedefaultseppunct}\relax
\EndOfBibitem
\bibitem[Borghi \latin{et~al.}(2014)Borghi, Ferretti, Nguyen, Dabo, and
  Marzari]{borghi_koopmans-compliant_2014}
Borghi,~G.; Ferretti,~A.; Nguyen,~N.~L.; Dabo,~I.; Marzari,~N.
  Koopmans-compliant functionals and their performance against reference
  molecular data. \emph{Phys. Rev. B} \textbf{2014}, \emph{90}, 075135\relax
\mciteBstWouldAddEndPuncttrue
\mciteSetBstMidEndSepPunct{\mcitedefaultmidpunct}
{\mcitedefaultendpunct}{\mcitedefaultseppunct}\relax
\EndOfBibitem
\bibitem[Ferretti \latin{et~al.}(2014)Ferretti, Dabo, Cococcioni, and
  Marzari]{ferretti_bridging_2014}
Ferretti,~A.; Dabo,~I.; Cococcioni,~M.; Marzari,~N. Bridging density-functional
  and many-body perturbation theory: {Orbital}-density dependence in
  electronic-structure functionals. \emph{Phys. Rev. B} \textbf{2014},
  \emph{89}, 195134\relax
\mciteBstWouldAddEndPuncttrue
\mciteSetBstMidEndSepPunct{\mcitedefaultmidpunct}
{\mcitedefaultendpunct}{\mcitedefaultseppunct}\relax
\EndOfBibitem
\bibitem[Gatti \latin{et~al.}(2007)Gatti, Olevano, Reining, and
  Tokatly]{gatti_transforming_2007}
Gatti,~M.; Olevano,~V.; Reining,~L.; Tokatly,~I.~V. Transforming {Nonlocality}
  into a {Frequency} {Dependence}: {A} {Shortcut} to {Spectroscopy}.
  \emph{Phys. Rev. Lett.} \textbf{2007}, \emph{99}, 057401\relax
\mciteBstWouldAddEndPuncttrue
\mciteSetBstMidEndSepPunct{\mcitedefaultmidpunct}
{\mcitedefaultendpunct}{\mcitedefaultseppunct}\relax
\EndOfBibitem
\bibitem[Vanzini \latin{et~al.}(2017)Vanzini, Reining, and
  Gatti]{vanzini_dynamical_2017}
Vanzini,~M.; Reining,~L.; Gatti,~M. Dynamical local connector approximation for
  electron addition and removal spectra. \emph{arXiv:1708.02450 [cond-mat]}
  \textbf{2017}, arXiv: 1708.02450\relax
\mciteBstWouldAddEndPuncttrue
\mciteSetBstMidEndSepPunct{\mcitedefaultmidpunct}
{\mcitedefaultendpunct}{\mcitedefaultseppunct}\relax
\EndOfBibitem
\bibitem[Vanzini \latin{et~al.}(2018)Vanzini, Reining, and
  Gatti]{vanzini_spectroscopy_2018}
Vanzini,~M.; Reining,~L.; Gatti,~M. Spectroscopy of the {Hubbard} dimer: the
  spectral potential. \emph{Eur. Phys. J. B} \textbf{2018}, \emph{91},
  192\relax
\mciteBstWouldAddEndPuncttrue
\mciteSetBstMidEndSepPunct{\mcitedefaultmidpunct}
{\mcitedefaultendpunct}{\mcitedefaultseppunct}\relax
\EndOfBibitem
\bibitem[Hedin(1965)]{hedin_new_1965}
Hedin,~L. New {Method} for {Calculating} the {One}-{Particle} {Green}'s
  {Function} with {Application} to the {Electron}-{Gas} {Problem}. \emph{Phys.
  Rev.} \textbf{1965}, \emph{139}, A796--A823\relax
\mciteBstWouldAddEndPuncttrue
\mciteSetBstMidEndSepPunct{\mcitedefaultmidpunct}
{\mcitedefaultendpunct}{\mcitedefaultseppunct}\relax
\EndOfBibitem
\bibitem[van Setten \latin{et~al.}(2015)van Setten, Caruso, Sharifzadeh, Ren,
  Scheffler, Liu, Lischner, Lin, Deslippe, Louie, Yang, Weigend, Neaton, Evers,
  and Rinke]{van_setten_gw100_2015}
van Setten,~M.~J.; Caruso,~F.; Sharifzadeh,~S.; Ren,~X.; Scheffler,~M.;
  Liu,~F.; Lischner,~J.; Lin,~L.; Deslippe,~J.~R.; Louie,~S.~G.; Yang,~C.;
  Weigend,~F.; Neaton,~J.~B.; Evers,~F.; Rinke,~P. {GW}100: {Benchmarking}
  {G}0W0 for {Molecular} {Systems}. \emph{J. Chem. Theory Comput.}
  \textbf{2015}, \emph{11}, 5665--5687\relax
\mciteBstWouldAddEndPuncttrue
\mciteSetBstMidEndSepPunct{\mcitedefaultmidpunct}
{\mcitedefaultendpunct}{\mcitedefaultseppunct}\relax
\EndOfBibitem
\bibitem[Nguyen \latin{et~al.}(2018)Nguyen, Colonna, Ferretti, and
  Marzari]{nguyen_koopmans-compliant_2017}
Nguyen,~N.~L.; Colonna,~N.; Ferretti,~A.; Marzari,~N. Koopmans-Compliant
  Spectral Functionals for Extended Systems. \emph{Phys. Rev. X} \textbf{2018},
  \emph{8}, 021051\relax
\mciteBstWouldAddEndPuncttrue
\mciteSetBstMidEndSepPunct{\mcitedefaultmidpunct}
{\mcitedefaultendpunct}{\mcitedefaultseppunct}\relax
\EndOfBibitem
\bibitem[Colonna \latin{et~al.}(2018)Colonna, Nguyen, Ferretti, and
  Marzari]{colonna_screening_2018}
Colonna,~N.; Nguyen,~N.~L.; Ferretti,~A.; Marzari,~N. Screening in
  {Orbital}-{Density}-{Dependent} {Functionals}. \emph{J. Chem. Theory Comput.}
  \textbf{2018}, \emph{14}, 2549--2557\relax
\mciteBstWouldAddEndPuncttrue
\mciteSetBstMidEndSepPunct{\mcitedefaultmidpunct}
{\mcitedefaultendpunct}{\mcitedefaultseppunct}\relax
\EndOfBibitem
\bibitem[Perdew and Zunger(1981)Perdew, and
  Zunger]{perdew_self-interaction_1981}
Perdew,~J.~P.; Zunger,~A. Self-interaction correction to density-functional
  approximations for many-electron systems. \emph{Phys. Rev. B} \textbf{1981},
  \emph{23}, 5048--5079\relax
\mciteBstWouldAddEndPuncttrue
\mciteSetBstMidEndSepPunct{\mcitedefaultmidpunct}
{\mcitedefaultendpunct}{\mcitedefaultseppunct}\relax
\EndOfBibitem
\bibitem[Cohen \latin{et~al.}(2008)Cohen, Mori-S{\'a}nchez, and
  Yang]{cohen_insights_2008}
Cohen,~A.~J.; Mori-S{\'a}nchez,~P.; Yang,~W. Insights into {Current}
  {Limitations} of {Density} {Functional} {Theory}. \emph{Science}
  \textbf{2008}, \emph{321}, 792--794\relax
\mciteBstWouldAddEndPuncttrue
\mciteSetBstMidEndSepPunct{\mcitedefaultmidpunct}
{\mcitedefaultendpunct}{\mcitedefaultseppunct}\relax
\EndOfBibitem
\bibitem[fhx()]{fhxc_pzsic}
Actually for the case of KIPZ one would need the second derivative of the
  PZ-SIC functional with respect to the orbital density, rather than the total
  density.\relax
\mciteBstWouldAddEndPunctfalse
\mciteSetBstMidEndSepPunct{\mcitedefaultmidpunct}
{}{\mcitedefaultseppunct}\relax
\EndOfBibitem
\bibitem[Giannozzi \latin{et~al.}(2017)Giannozzi, Andreussi, Brumme, Bunau,
  Nardelli, Calandra, Car, Cavazzoni, {D Ceresoli}, Cococcioni, Colonna,
  Carnimeo, Corso, Gironcoli, Delugas, Jr, {A Ferretti}, Floris, Fratesi,
  Fugallo, Gebauer, Gerstmann, Giustino, Gorni, Jia, Kawamura, {H-Y Ko},
  Kokalj, K{\"u}{\c c}{\"u}kbenli, Lazzeri, Marsili, Marzari, Mauri, Nguyen,
  Nguyen, {A Otero-de-la-Roza}, Paulatto, Ponc{\'e}, Rocca, Sabatini, Santra,
  Schlipf, Seitsonen, Smogunov, {I Timrov}, Thonhauser, Umari, Vast, Wu, and
  Baroni]{giannozzi_advanced_2017}
Giannozzi,~P. \latin{et~al.}  Advanced capabilities for materials modelling
  with {Q} uantum {ESPRESSO}. \emph{J. Phys.: Condens. Matter} \textbf{2017},
  \emph{29}, 465901\relax
\mciteBstWouldAddEndPuncttrue
\mciteSetBstMidEndSepPunct{\mcitedefaultmidpunct}
{\mcitedefaultendpunct}{\mcitedefaultseppunct}\relax
\EndOfBibitem
\bibitem[Gonze \latin{et~al.}(2016)Gonze, Jollet, Abreu~Araujo, Adams, Amadon,
  Applencourt, Audouze, Beuken, Bieder, Bokhanchuk, Bousquet, Bruneval,
  Caliste, C{\^o}t{\'e}, Dahm, Da~Pieve, Delaveau, Di~Gennaro, Dorado, Espejo,
  Geneste, Genovese, Gerossier, Giantomassi, Gillet, Hamann, He, Jomard,
  Laflamme~Janssen, Le~Roux, Levitt, Lherbier, Liu, Luka{\v c}evi{\'c}, Martin,
  Martins, Oliveira, Ponc{\'e}, Pouillon, Rangel, Rignanese, Romero, Rousseau,
  Rubel, Shukri, Stankovski, Torrent, Van~Setten, Van~Troeye, Verstraete,
  Waroquiers, Wiktor, Xu, Zhou, and Zwanziger]{gonze_recent_2016}
Gonze,~X. \latin{et~al.}  Recent developments in the {ABINIT} software package.
  \emph{Computer Physics Communications} \textbf{2016}, \emph{205},
  106--131\relax
\mciteBstWouldAddEndPuncttrue
\mciteSetBstMidEndSepPunct{\mcitedefaultmidpunct}
{\mcitedefaultendpunct}{\mcitedefaultseppunct}\relax
\EndOfBibitem
\bibitem[Nguyen \latin{et~al.}(2015)Nguyen, Borghi, Ferretti, Dabo, and
  Marzari]{nguyen_first-principles_2015}
Nguyen,~N.~L.; Borghi,~G.; Ferretti,~A.; Dabo,~I.; Marzari,~N.
  First-{Principles} {Photoemission} {Spectroscopy} and {Orbital} {Tomography}
  in {Molecules} from {Koopmans}-{Compliant} {Functionals}. \emph{Phys. Rev.
  Lett.} \textbf{2015}, \emph{114}, 166405\relax
\mciteBstWouldAddEndPuncttrue
\mciteSetBstMidEndSepPunct{\mcitedefaultmidpunct}
{\mcitedefaultendpunct}{\mcitedefaultseppunct}\relax
\EndOfBibitem
\bibitem[Lehtola and J{\'o}nsson(2014)Lehtola, and
  J{\'o}nsson]{lehtola_variational_2014}
Lehtola,~S.; J{\'o}nsson,~H. Variational, {Self}-{Consistent} {Implementation}
  of the {Perdew}{\textendash}{Zunger} {Self}-{Interaction} {Correction} with
  {Complex} {Optimal} {Orbitals}. \emph{J. Chem. Theory Comput.} \textbf{2014},
  \emph{10}, 5324--5337\relax
\mciteBstWouldAddEndPuncttrue
\mciteSetBstMidEndSepPunct{\mcitedefaultmidpunct}
{\mcitedefaultendpunct}{\mcitedefaultseppunct}\relax
\EndOfBibitem
\bibitem[Vydrov \latin{et~al.}(2007)Vydrov, Scuseria, and
  Perdew]{vydrov_tests_2007}
Vydrov,~O.~A.; Scuseria,~G.~E.; Perdew,~J.~P. Tests of functionals for systems
  with fractional electron number. \emph{J. Chem. Phys.} \textbf{2007},
  \emph{126}, 154109\relax
\mciteBstWouldAddEndPuncttrue
\mciteSetBstMidEndSepPunct{\mcitedefaultmidpunct}
{\mcitedefaultendpunct}{\mcitedefaultseppunct}\relax
\EndOfBibitem
\bibitem[Pederson \latin{et~al.}(1984)Pederson, Heaton, and
  Lin]{pederson_localdensity_1984}
Pederson,~M.~R.; Heaton,~R.~A.; Lin,~C.~C. Local-density
  {Hartree}{\textendash}{Fock} theory of electronic states of molecules with
  self-interaction correction. \emph{J. Chem. Phys.} \textbf{1984}, \emph{80},
  1972--1975\relax
\mciteBstWouldAddEndPuncttrue
\mciteSetBstMidEndSepPunct{\mcitedefaultmidpunct}
{\mcitedefaultendpunct}{\mcitedefaultseppunct}\relax
\EndOfBibitem
\bibitem[Stengel and Spaldin(2008)Stengel, and
  Spaldin]{stengel_self-interaction_2008}
Stengel,~M.; Spaldin,~N.~A. Self-interaction correction with {Wannier}
  functions. \emph{Phys. Rev. B} \textbf{2008}, \emph{77}, 155106\relax
\mciteBstWouldAddEndPuncttrue
\mciteSetBstMidEndSepPunct{\mcitedefaultmidpunct}
{\mcitedefaultendpunct}{\mcitedefaultseppunct}\relax
\EndOfBibitem
\bibitem[Hofmann \latin{et~al.}(2012)Hofmann, Kl{\"u}pfel, Kl{\"u}pfel, and
  K{\"u}mmel]{hofmann_using_2012}
Hofmann,~D.; Kl{\"u}pfel,~S.; Kl{\"u}pfel,~P.; K{\"u}mmel,~S. Using complex
  degrees of freedom in the {Kohn}-{Sham} self-interaction correction.
  \emph{Phys. Rev. A} \textbf{2012}, \emph{85}, 062514\relax
\mciteBstWouldAddEndPuncttrue
\mciteSetBstMidEndSepPunct{\mcitedefaultmidpunct}
{\mcitedefaultendpunct}{\mcitedefaultseppunct}\relax
\EndOfBibitem
\bibitem[Pederson and Lin(1988)Pederson, and Lin]{pederson_localized_1988}
Pederson,~M.~R.; Lin,~C.~C. Localized and canonical atomic orbitals in
  self-interaction corrected local density functional approximation. \emph{J.
  Chem. Phys.} \textbf{1988}, \emph{88}, 1807--1817\relax
\mciteBstWouldAddEndPuncttrue
\mciteSetBstMidEndSepPunct{\mcitedefaultmidpunct}
{\mcitedefaultendpunct}{\mcitedefaultseppunct}\relax
\EndOfBibitem
\bibitem[Edmiston and Ruedenberg(1963)Edmiston, and
  Ruedenberg]{edmiston_localized_1963}
Edmiston,~C.; Ruedenberg,~K. Localized Atomic and Molecular Orbitals.
  \emph{Rev. Mod. Phys.} \textbf{1963}, \emph{35}, 457--464\relax
\mciteBstWouldAddEndPuncttrue
\mciteSetBstMidEndSepPunct{\mcitedefaultmidpunct}
{\mcitedefaultendpunct}{\mcitedefaultseppunct}\relax
\EndOfBibitem
\bibitem[Boys(1960)]{boys_construction_1960}
Boys,~S.~F. Construction of {Some} {Molecular} {Orbitals} to {Be}
  {Approximately} {Invariant} for {Changes} from {One} {Molecule} to {Another}.
  \emph{Rev. Mod. Phys.} \textbf{1960}, \emph{32}, 296--299\relax
\mciteBstWouldAddEndPuncttrue
\mciteSetBstMidEndSepPunct{\mcitedefaultmidpunct}
{\mcitedefaultendpunct}{\mcitedefaultseppunct}\relax
\EndOfBibitem
\bibitem[Foster and Boys(1960)Foster, and Boys]{foster_canonical_1960}
Foster,~J.~M.; Boys,~S.~F. Canonical {Configurational} {Interaction}
  {Procedure}. \emph{Rev. Mod. Phys.} \textbf{1960}, \emph{32}, 300--302\relax
\mciteBstWouldAddEndPuncttrue
\mciteSetBstMidEndSepPunct{\mcitedefaultmidpunct}
{\mcitedefaultendpunct}{\mcitedefaultseppunct}\relax
\EndOfBibitem
\bibitem[Marzari \latin{et~al.}(2012)Marzari, Mostofi, Yates, Souza, and
  Vanderbilt]{marzari_maximally_2012}
Marzari,~N.; Mostofi,~A.~A.; Yates,~J.~R.; Souza,~I.; Vanderbilt,~D. Maximally
  localized {Wannier} functions: {Theory} and applications. \emph{Rev. Mod.
  Phys.} \textbf{2012}, \emph{84}, 1419--1475\relax
\mciteBstWouldAddEndPuncttrue
\mciteSetBstMidEndSepPunct{\mcitedefaultmidpunct}
{\mcitedefaultendpunct}{\mcitedefaultseppunct}\relax
\EndOfBibitem
\bibitem[ki_()]{ki_invariance}
KI functional at integer occupations is invariant under unitary rotations; a
  representation is choosen~\cite{borghi_koopmans-compliant_2014} by taking it
  as the limit of the KIPZ functional when the PZ correction tends to zero (as
  discussed in Section 4.1).\relax
\mciteBstWouldAddEndPunctfalse
\mciteSetBstMidEndSepPunct{\mcitedefaultmidpunct}
{}{\mcitedefaultseppunct}\relax
\EndOfBibitem
\bibitem[Szabo and Ostlund(1996)Szabo, and Ostlund]{szabo_modern_1996}
Szabo,~A.; Ostlund,~N.~S. \emph{Modern {Quantum} {Chemistry}: {Introduction} to
  {Advanced} {Electronic} {Structure} {Theory}}, revised. edizione ed.; Dover
  Pubns: Mineola, N.Y., 1996\relax
\mciteBstWouldAddEndPuncttrue
\mciteSetBstMidEndSepPunct{\mcitedefaultmidpunct}
{\mcitedefaultendpunct}{\mcitedefaultseppunct}\relax
\EndOfBibitem
\bibitem[Gygi and Baldereschi(1989)Gygi, and
  Baldereschi]{gygi_quasiparticle_1989}
Gygi,~F.; Baldereschi,~A. Quasiparticle energies in semiconductors:
  {Self}-energy correction to the local-density approximation. \emph{Phys. Rev.
  Lett.} \textbf{1989}, \emph{62}, 2160--2163\relax
\mciteBstWouldAddEndPuncttrue
\mciteSetBstMidEndSepPunct{\mcitedefaultmidpunct}
{\mcitedefaultendpunct}{\mcitedefaultseppunct}\relax
\EndOfBibitem
\bibitem[Kang and Hybertsen(2010)Kang, and Hybertsen]{kang_enhanced_2010}
Kang,~W.; Hybertsen,~M.~S. Enhanced static approximation to the electron
  self-energy operator for efficient calculation of quasiparticle energies.
  \emph{Phys. Rev. B} \textbf{2010}, \emph{82}, 195108\relax
\mciteBstWouldAddEndPuncttrue
\mciteSetBstMidEndSepPunct{\mcitedefaultmidpunct}
{\mcitedefaultendpunct}{\mcitedefaultseppunct}\relax
\EndOfBibitem
\bibitem[Hybertsen and Louie(1987)Hybertsen, and Louie]{hybertsen_ab_1987}
Hybertsen,~M.~S.; Louie,~S.~G. Ab initio static dielectric matrices from the
  density-functional approach. {I}. {Formulation} and application to
  semiconductors and insulators. \emph{Phys. Rev. B} \textbf{1987}, \emph{35},
  5585--5601\relax
\mciteBstWouldAddEndPuncttrue
\mciteSetBstMidEndSepPunct{\mcitedefaultmidpunct}
{\mcitedefaultendpunct}{\mcitedefaultseppunct}\relax
\EndOfBibitem
\bibitem[Del~Sole \latin{et~al.}(1994)Del~Sole, Reining, and
  Godby]{del_sole_gwgamma_1994}
Del~Sole,~R.; Reining,~L.; Godby,~R.~W. $GW\Gamma$ approximation for electron
  self-energies in semiconductors and insulators. \emph{Phys. Rev. B}
  \textbf{1994}, \emph{49}, 8024--8028\relax
\mciteBstWouldAddEndPuncttrue
\mciteSetBstMidEndSepPunct{\mcitedefaultmidpunct}
{\mcitedefaultendpunct}{\mcitedefaultseppunct}\relax
\EndOfBibitem
\bibitem[Bruneval \latin{et~al.}(2005)Bruneval, Sottile, Olevano, Del~Sole, and
  Reining]{bruneval_many-body_2005}
Bruneval,~F.; Sottile,~F.; Olevano,~V.; Del~Sole,~R.; Reining,~L. Many-{Body}
  {Perturbation} {Theory} {Using} the {Density}-{Functional} {Concept}:
  {Beyond} the \${GW}\$ {Approximation}. \emph{Phys. Rev. Lett.} \textbf{2005},
  \emph{94}, 186402\relax
\mciteBstWouldAddEndPuncttrue
\mciteSetBstMidEndSepPunct{\mcitedefaultmidpunct}
{\mcitedefaultendpunct}{\mcitedefaultseppunct}\relax
\EndOfBibitem
\bibitem[Gross and Kohn(1985)Gross, and Kohn]{gross_local_1985}
Gross,~E. K.~U.; Kohn,~W. Local density-functional theory of
  frequency-dependent linear response. \emph{Phys. Rev. Lett.} \textbf{1985},
  \emph{55}, 2850--2852\relax
\mciteBstWouldAddEndPuncttrue
\mciteSetBstMidEndSepPunct{\mcitedefaultmidpunct}
{\mcitedefaultendpunct}{\mcitedefaultseppunct}\relax
\EndOfBibitem
\bibitem[Petersilka \latin{et~al.}(1996)Petersilka, Gossmann, and
  Gross]{petersilka_excitation_1996}
Petersilka,~M.; Gossmann,~U.~J.; Gross,~E. K.~U. Excitation {Energies} from
  {Time}-{Dependent} {Density}-{Functional} {Theory}. \emph{Phys. Rev. Lett.}
  \textbf{1996}, \emph{76}, 1212--1215\relax
\mciteBstWouldAddEndPuncttrue
\mciteSetBstMidEndSepPunct{\mcitedefaultmidpunct}
{\mcitedefaultendpunct}{\mcitedefaultseppunct}\relax
\EndOfBibitem
\bibitem[Hybertsen and Louie(1986)Hybertsen, and
  Louie]{hybertsen_electron_1986}
Hybertsen,~M.~S.; Louie,~S.~G. Electron correlation in semiconductors and
  insulators: {Band} gaps and quasiparticle energies. \emph{Phys. Rev. B}
  \textbf{1986}, \emph{34}, 5390--5413\relax
\mciteBstWouldAddEndPuncttrue
\mciteSetBstMidEndSepPunct{\mcitedefaultmidpunct}
{\mcitedefaultendpunct}{\mcitedefaultseppunct}\relax
\EndOfBibitem
\bibitem[cas()]{casida_vxc_note}
A rigorous justification for such a starting point is provided by the fact that
  the exact KS exchange-correlation potential is the variationally best local
  approximation to the exchange-correlation
  self-energy~\cite{casida_generalization_1995}\relax
\mciteBstWouldAddEndPuncttrue
\mciteSetBstMidEndSepPunct{\mcitedefaultmidpunct}
{\mcitedefaultendpunct}{\mcitedefaultseppunct}\relax
\EndOfBibitem
\bibitem[Giannozzi \latin{et~al.}(2009)Giannozzi, Baroni, Bonini, Calandra,
  Car, Cavazzoni, Ceresoli, Chiarotti, Cococcioni, Dabo, Corso, Gironcoli,
  Fabris, Fratesi, Gebauer, Gerstmann, Gougoussis, Kokalj, Lazzeri,
  Martin-Samos, Marzari, Mauri, Mazzarello, Paolini, Pasquarello, Paulatto,
  Sbraccia, Scandolo, Sclauzero, Seitsonen, Smogunov, Umari, and
  Wentzcovitch]{giannozzi_quantum_2009}
Giannozzi,~P. \latin{et~al.}  {QUANTUM} {ESPRESSO}: a modular and open-source
  software project for quantum simulations of materials. \emph{J. Phys.:
  Conden. Matter} \textbf{2009}, \emph{21}, 395502\relax
\mciteBstWouldAddEndPuncttrue
\mciteSetBstMidEndSepPunct{\mcitedefaultmidpunct}
{\mcitedefaultendpunct}{\mcitedefaultseppunct}\relax
\EndOfBibitem
\bibitem[Li and Dabo(2011)Li, and Dabo]{li_electronic_2011}
Li,~Y.; Dabo,~I. Electronic levels and electrical response of periodic
  molecular structures from plane-wave orbital-dependent calculations.
  \emph{Phys. Rev. B} \textbf{2011}, \emph{84}, 155127\relax
\mciteBstWouldAddEndPuncttrue
\mciteSetBstMidEndSepPunct{\mcitedefaultmidpunct}
{\mcitedefaultendpunct}{\mcitedefaultseppunct}\relax
\EndOfBibitem
\bibitem[Martyna and Tuckerman(1999)Martyna, and
  Tuckerman]{martyna_reciprocal_1999}
Martyna,~G.~J.; Tuckerman,~M.~E. A reciprocal space based method for treating
  long range interactions in ab initio and force-field-based calculations in
  clusters. \emph{J. Chem. Phys.} \textbf{1999}, \emph{110}, 2810--2821\relax
\mciteBstWouldAddEndPuncttrue
\mciteSetBstMidEndSepPunct{\mcitedefaultmidpunct}
{\mcitedefaultendpunct}{\mcitedefaultseppunct}\relax
\EndOfBibitem
\bibitem[Hamann(2013)]{hamann_optimized_2013}
Hamann,~D.~R. Optimized norm-conserving {Vanderbilt} pseudopotentials.
  \emph{Phys. Rev. B} \textbf{2013}, \emph{88}, 085117\relax
\mciteBstWouldAddEndPuncttrue
\mciteSetBstMidEndSepPunct{\mcitedefaultmidpunct}
{\mcitedefaultendpunct}{\mcitedefaultseppunct}\relax
\EndOfBibitem
\bibitem[Schlipf and Gygi(2015)Schlipf, and Gygi]{schlipf_optimization_2015}
Schlipf,~M.; Gygi,~F. Optimization algorithm for the generation of {ONCV}
  pseudopotentials. \emph{Comput. Phys. Commun.} \textbf{2015}, \emph{196},
  36--44\relax
\mciteBstWouldAddEndPuncttrue
\mciteSetBstMidEndSepPunct{\mcitedefaultmidpunct}
{\mcitedefaultendpunct}{\mcitedefaultseppunct}\relax
\EndOfBibitem
\bibitem[ONC(2017)]{ONCV_website}
{SG15} {ONCV} {Potentials}. 2017;
  \url{http://www.quantum-simulation.org/potentials/sg15_oncv/}\relax
\mciteBstWouldAddEndPuncttrue
\mciteSetBstMidEndSepPunct{\mcitedefaultmidpunct}
{\mcitedefaultendpunct}{\mcitedefaultseppunct}\relax
\EndOfBibitem
\bibitem[Perdew \latin{et~al.}(1996)Perdew, Burke, and
  Ernzerhof]{perdew_generalized_1996}
Perdew,~J.~P.; Burke,~K.; Ernzerhof,~M. Generalized {Gradient} {Approximation}
  {Made} {Simple}. \emph{Phys. Rev. Lett.} \textbf{1996}, \emph{77},
  3865--3868\relax
\mciteBstWouldAddEndPuncttrue
\mciteSetBstMidEndSepPunct{\mcitedefaultmidpunct}
{\mcitedefaultendpunct}{\mcitedefaultseppunct}\relax
\EndOfBibitem
\bibitem[Marzari \latin{et~al.}(1997)Marzari, Vanderbilt, and
  Payne]{marzari_ensemble_1997}
Marzari,~N.; Vanderbilt,~D.; Payne,~M.~C. Ensemble {Density}-{Functional}
  {Theory} for {Ab} {Initio} {Molecular} {Dynamics} of {Metals} and
  {Finite}-{Temperature} {Insulators}. \emph{Phys. Rev. Lett.} \textbf{1997},
  \emph{79}, 1337--1340\relax
\mciteBstWouldAddEndPuncttrue
\mciteSetBstMidEndSepPunct{\mcitedefaultmidpunct}
{\mcitedefaultendpunct}{\mcitedefaultseppunct}\relax
\EndOfBibitem
\bibitem[Borghi \latin{et~al.}(2015)Borghi, Park, Nguyen, Ferretti, and
  Marzari]{borghi_variational_2015}
Borghi,~G.; Park,~C.-H.; Nguyen,~N.~L.; Ferretti,~A.; Marzari,~N. Variational
  minimization of orbital-density-dependent functionals. \emph{Phys. Rev. B}
  \textbf{2015}, \emph{91}, 155112\relax
\mciteBstWouldAddEndPuncttrue
\mciteSetBstMidEndSepPunct{\mcitedefaultmidpunct}
{\mcitedefaultendpunct}{\mcitedefaultseppunct}\relax
\EndOfBibitem
\bibitem[ip_()]{ip_ehomo_note}
A direct identification of $-\varepsilon_{i}$ with the IP is possible thanks to
  the use of counter-charge corrections that enforce the correct asymptotic
  limit of the potential, that is $v(\mathbf{r})=0$ (vacuum level) when
  $\mathbf{r}\rightarrow \infty$ (i.e. far from the system)\relax
\mciteBstWouldAddEndPuncttrue
\mciteSetBstMidEndSepPunct{\mcitedefaultmidpunct}
{\mcitedefaultendpunct}{\mcitedefaultseppunct}\relax
\EndOfBibitem
\bibitem[Krause \latin{et~al.}(2015)Krause, Harding, and
  Klopper]{krause_coupled-cluster_2015}
Krause,~K.; Harding,~M.~E.; Klopper,~W. Coupled-cluster reference values for
  the {GW}27 and {GW}100 test sets for the assessment of {GW} methods.
  \emph{Molecular Physics} \textbf{2015}, \emph{113}, 1952--1960\relax
\mciteBstWouldAddEndPuncttrue
\mciteSetBstMidEndSepPunct{\mcitedefaultmidpunct}
{\mcitedefaultendpunct}{\mcitedefaultseppunct}\relax
\EndOfBibitem
\bibitem[Pela \latin{et~al.}(2018)Pela, Gulans, and Draxl]{pela_lda-1/2_2018}
Pela,~R.~R.; Gulans,~A.; Draxl,~C. The {LDA}-1/2 method applied to atoms and
  molecules. \emph{arXiv:1805.09705 [cond-mat, physics:physics]} \textbf{2018},
  arXiv: 1805.09705\relax
\mciteBstWouldAddEndPuncttrue
\mciteSetBstMidEndSepPunct{\mcitedefaultmidpunct}
{\mcitedefaultendpunct}{\mcitedefaultseppunct}\relax
\EndOfBibitem
\bibitem[Brawand \latin{et~al.}(2017)Brawand, Govoni, V{\"o}r{\"o}s, and
  Galli]{brawand_performance_2017}
Brawand,~N.~P.; Govoni,~M.; V{\"o}r{\"o}s,~M.; Galli,~G. Performance and
  {Self}-{Consistency} of the {Generalized} {Dielectric} {Dependent} {Hybrid}
  {Functional}. \emph{J. Chem. Theory Comput.} \textbf{2017}, \relax
\mciteBstWouldAddEndPunctfalse
\mciteSetBstMidEndSepPunct{\mcitedefaultmidpunct}
{}{\mcitedefaultseppunct}\relax
\EndOfBibitem
\bibitem[Govoni and Galli(2018)Govoni, and Galli]{govoni_gw100_2018}
Govoni,~M.; Galli,~G. {GW}100: {Comparison} of {Methods} and {Accuracy} of
  {Results} {Obtained} with the {WEST} {Code}. \emph{J. Chem. Theory Comput.}
  \textbf{2018}, \emph{14}, 1895--1909\relax
\mciteBstWouldAddEndPuncttrue
\mciteSetBstMidEndSepPunct{\mcitedefaultmidpunct}
{\mcitedefaultendpunct}{\mcitedefaultseppunct}\relax
\EndOfBibitem
\bibitem[Caruso \latin{et~al.}(2016)Caruso, Dauth, van Setten, and
  Rinke]{caruso_benchmark_2016}
Caruso,~F.; Dauth,~M.; van Setten,~M.~J.; Rinke,~P. Benchmark of {GW}
  {Approaches} for the {GW}100 {Test} {Set}. \emph{J. Chem. Theory Comput.}
  \textbf{2016}, \emph{12}, 5076--5087\relax
\mciteBstWouldAddEndPuncttrue
\mciteSetBstMidEndSepPunct{\mcitedefaultmidpunct}
{\mcitedefaultendpunct}{\mcitedefaultseppunct}\relax
\EndOfBibitem
\bibitem[Ferreira \latin{et~al.}(2008)Ferreira, Marques, and
  Teles]{ferreira_approximation_2008}
Ferreira,~L.~G.; Marques,~M.; Teles,~L.~K. Approximation to density functional
  theory for the calculation of band gaps of semiconductors. \emph{Phys. Rev.
  B} \textbf{2008}, \emph{78}, 125116\relax
\mciteBstWouldAddEndPuncttrue
\mciteSetBstMidEndSepPunct{\mcitedefaultmidpunct}
{\mcitedefaultendpunct}{\mcitedefaultseppunct}\relax
\EndOfBibitem
\bibitem[Skone \latin{et~al.}(2014)Skone, Govoni, and
  Galli]{skone_self-consistent_2014}
Skone,~J.~H.; Govoni,~M.; Galli,~G. Self-consistent hybrid functional for
  condensed systems. \emph{Phys. Rev. B} \textbf{2014}, \emph{89}, 195112\relax
\mciteBstWouldAddEndPuncttrue
\mciteSetBstMidEndSepPunct{\mcitedefaultmidpunct}
{\mcitedefaultendpunct}{\mcitedefaultseppunct}\relax
\EndOfBibitem
\bibitem[Skone \latin{et~al.}(2016)Skone, Govoni, and
  Galli]{skone_nonempirical_2016}
Skone,~J.~H.; Govoni,~M.; Galli,~G. Nonempirical range-separated hybrid
  functionals for solids and molecules. \emph{Phys. Rev. B} \textbf{2016},
  \emph{93}, 235106\relax
\mciteBstWouldAddEndPuncttrue
\mciteSetBstMidEndSepPunct{\mcitedefaultmidpunct}
{\mcitedefaultendpunct}{\mcitedefaultseppunct}\relax
\EndOfBibitem
\bibitem[Brawand \latin{et~al.}(2016)Brawand, V{\"o}r{\"o}s, Govoni, and
  Galli]{brawand_generalization_2016}
Brawand,~N.~P.; V{\"o}r{\"o}s,~M.; Govoni,~M.; Galli,~G. Generalization of
  {Dielectric}-{Dependent} {Hybrid} {Functionals} to {Finite} {Systems}.
  \emph{Phys. Rev. X} \textbf{2016}, \emph{6}, 041002\relax
\mciteBstWouldAddEndPuncttrue
\mciteSetBstMidEndSepPunct{\mcitedefaultmidpunct}
{\mcitedefaultendpunct}{\mcitedefaultseppunct}\relax
\EndOfBibitem
\bibitem[Shapley and Chong(2001)Shapley, and Chong]{shapley_pw86pw91_2001}
Shapley,~W.~A.; Chong,~D.~P. {PW}86{\textendash}{PW}91 density functional
  calculation of vertical ionization potentials: Some implications for
  present-day functionals. \emph{Int. J. Quant. Chem.} \textbf{2001},
  \emph{81}, 34--52\relax
\mciteBstWouldAddEndPuncttrue
\mciteSetBstMidEndSepPunct{\mcitedefaultmidpunct}
{\mcitedefaultendpunct}{\mcitedefaultseppunct}\relax
\EndOfBibitem
\end{mcitethebibliography}

\providecommand{\latin}[1]{#1}
\providecommand*\mcitethebibliography{\thebibliography}
\csname @ifundefined\endcsname{endmcitethebibliography}
  {\let\endmcitethebibliography\endthebibliography}{}

\end{document}